\documentclass[aps,twocolumn,floatfix,showpacs,amsmath,amsfonts]{revtex4-1}

\usepackage{graphicx}
\usepackage{bm}
\usepackage{units}
\usepackage{color}
\usepackage{multirow}
\usepackage{upgreek}

\begin{document}

\title{The even exciton series in Cu$_2$O}

\author{Frank Schweiner}
\author{J\"org Main}
\author{G\"unter Wunner}
\affiliation{Institut f\"ur Theoretische Physik 1, Universit\"at Stuttgart,
  70550 Stuttgart, Germany}
\author{Christoph Uihlein}
\affiliation{Experimentelle Physik 2, Technische Universit\"at Dortmund, 44221 Dortmund, Germany}
\date{\today}

\begin{abstract}
Recent investigations of excitonic absorption spectra in cuprous oxide
($\mathrm{Cu_{2}O}$) have shown that it is indispensable to account for the
complex valence band structure in the theory of excitons. 
In $\mathrm{Cu_{2}O}$ parity is a good quantum number and thus the exciton 
spectrum falls into two parts: The dipole-active exciton states 
of negative parity and odd angular momentum, which can be observed 
in one-photon absorption ($\Gamma_4^-$ symmetry) and the exciton states of 
positive parity and even angular momentum, which can be observed in 
two-photon absorption ($\Gamma_5^+$ symmetry). 
The unexpected observation of 
$D$ excitons in two-photon absorption has given first evidence 
that the dispersion properties of the $\Gamma_5^+$ orbital valence band is 
giving rise to a coupling of the yellow and green exciton series. 
However, a first theoretical treatment by Ch.~Uihlein~\emph{et al.} [Phys.~Rev.~B~\textbf{23}, 2731 (1981)]
was based on a simplified spherical model. 
The observation of $F$ excitons in one-photon 
absorption is a further proof of a coupling between yellow and green 
exciton states. Detailed investigations on the fine structure splitting 
of the $F$ exciton by F.~Schweiner~\emph{et al.} [Phys.~Rev.~B~\textbf{93}, 195203 (2016)]
have proved the importance of a more realistic theoretical 
treatment including terms with cubic symmetry. 
In this paper we show that the even and odd parity exciton system can be consistently 
described within the same theoretical approach. However, the Hamiltonian of 
the even parity system needs, in comparison to the odd exciton case,
modifications to account for the very small radius of the yellow and green 
$1S$ exciton. In the presented treatment we take special care of the
central-cell corrections, which comprise a reduced 
screening of the Coulomb potential at distances comparable to the polaron 
radius, the exchange interaction being responsible 
for the exciton splitting into ortho and para states,
and the inclusion of terms in the fourth 
power of $p$ in the kinetic energy being consistent with $O_{\mathrm{h}}$ symmetry.
Since the yellow $1S$ exciton state is coupled to all other states of positive parity, 
we show how the central-cell corrections affect the whole even exciton series.
The close resonance of the $1S$ green exciton with states of the 
yellow exciton series has a strong impact on the energies and 
oscillator strengths of all implied states. 
The consistency between theory and 
experiment with respect to energies and oscillator strengths for the even and odd exciton 
system in $\mathrm{Cu_{2}O}$ is a convincing proof for the validity of the applied theory.
\end{abstract}


\maketitle

\section{Introduction~\label{sec:Introduction}}

Excitons are the quanta of fundamental optical excitations in both 
insulators and semiconductors in the visible and ultraviolet spectrum 
of light. The Coulomb interaction between electron and hole leads to 
a hydrogen-like series of excitonic states~\cite{TOE}. 
Cuprous oxide $\left(\mathrm{Cu_{2}O}\right)$ 
is a prime example where one can even identify four different excitonic 
series (yellow, green, blue, and violet) being related to the two 
topmost valence bands and the two lowest conduction bands~\cite{GRE}.  
Recently, the yellow series could be followed up to a spectacular 
high principal quantum number of $n=25$~\cite{GRE}. 
This outstanding experiment has launched 
the new field of research of giant Rydberg excitons
and led to a variety of new theoretical and experimental
investigations on the topic of excitons in 
$\mathrm{Cu_{2}O}$~\cite{28,QC,175,75,76,50,80,100,125,78,79,150,74,77,225}.

$\mathrm{Cu_{2}O}$ has octahedral symmetry $O_{\mathrm{h}}$ so that
the symmetry of the bands can be assigned by the 
irreducible representations $\Gamma_{i}^{\pm}$
of $O_{\mathrm{h}}$. 
The yellow and green exciton series share the 
same threefold degenerate $\Gamma_5^+$ orbital valence band state. This state splits due 
to spin-orbit interaction into an upper twofold degenerate $\Gamma_7^+$ valence band 
(yellow series) and a lower fourfold degenerate $\Gamma_8^+$ valence band (green series). 
The band structure of both bands is essentially determined by the anisotropic 
dispersion properties of the orbital state. The threefold degeneracy of the 
orbital state is lifted as soon as a non-zero $k$ vector gets involved, with 
new eigenvectors depending on the orientation of $\boldsymbol{k}$. A consequence of the 
splitting of the orbital state is a partial quenching of the spin-orbit interaction. 
This $\boldsymbol{k}$ dependent quenching is not only responsible for a remarkable non-parabolicity 
of the two top valence bands but leads likewise to a $\boldsymbol{k}$ dependent mixing of the 
$\Gamma_7^+$ and $\Gamma_8^+$ Bloch states and can thus cause a mixing of the yellow and green 
exciton series. A mixing of both series is favored by the large Rydberg 
energy of approximately $100\,\mathrm{meV}$, a corresponding large exciton extension 
in $k$ space and the small spin-orbit splitting of only $130\,\mathrm{meV}$.

A Hamiltonian that is able to cope with a coupled system of yellow and green 
excitons must take explicit care of the dispersion properties of the orbital 
valence band state and has to include the spin-orbit interaction. Such a kind of 
Hamiltonian was first introduced by Uihlein~\emph{et~al.}~\cite{7} for explaining the unexpected 
fine structure splitting observed in the two-photon absorption spectrum of $\mathrm{Cu_{2}O}$. 
They used a simplified spherical dispersion model for the $\Gamma_5^+$ orbital valence 
band with an identical splitting into longitudinal and transverse states independent 
of the orientation of $\boldsymbol{k}$. This simplification had the appealing advantage that the 
total angular momentum remains a good quantum number so that the exciton problem could 
be reduced to calculate the eigenvalues of a system of coupled radial wave functions. 
A problem in their paper is the incorrect notation of the $1S$ green and $2S$ yellow excitons 
states. Both notations need to be exchanged to be consistent with their calculations. 
Although the spherical model can explain many details of the experimental findings, one 
has to be aware of its limitations. A more realistic Hamiltonian being compliant with 
the real band structure by including terms of cubic symmetry has already proved its 
validity by explaining the puzzling fine structure of the odd parity states in $\mathrm{Cu_{2}O}$~\cite{100}. 
The intention of this paper is to show that the same kind of Hamiltonian can 
likewise describe the fine splitting of the even parity excitons.

However, when comparing the even parity and odd parity exciton systems, it is obvious 
that the even exciton system is a much more challenging problem. One reason 
for this is the close resonance of the green $1S$ exciton with the even parity 
states of the yellow series with principal quantum number $n\geq 2$. This requires a 
very careful calculation of the binding energy of the green $1S$ exciton. 
Furthermore, the binding energy of the yellow $1S$ exciton is much larger than expected from a 
simple hydrogen like series, inter alia, due to a less effective screening of the 
Coulomb potential at distances comparable to the polaron radius. 
Moreover, a breakdown of the electronic screening is expected at even much shorter distances, but a 
proper treatment is exceeding the limits of the continuum approximation. Hence, we
introduce in this paper a $\delta$-function like central cell correction term that 
should account for all kinds of short range perturbations affecting the immediate 
neighborhood of the central cell. The magnitude of this term is treated as a free 
parameter that can be adjusted to the experimental findings. It is important to 
note that a change of this parameter leads to a significant shift of the green 
$1S$ exciton with respect to the higher order states of the yellow series and has 
therefore a high impact on the energies and the compositions of the resulting 
coupled exciton states. Taking this in mind it is fundamental that one can 
likewise achieve a match to the relative oscillator strengths of the involved states.

Dealing with the even parity system of $\mathrm{Cu_{2}O}$ is also confronting us with the problem 
of a proper treatment of the $1S$ exciton with respect to its very small radius since
a small exciton radius means a large extension of the exciton in $k$ space. The 
challenge is therefore to meet the band structure of the valence band in a much 
larger vicinity of the $\Gamma$ point. For coping with this situation, we include
in the kinetic energy of the hole all terms in the fourth power of $p$ being 
compliant with the octahedral symmetry of $\mathrm{Cu_{2}O}$. The parameters of these terms 
are carefully adjusted to get a best fit to the band structure in the part of 
the $k$ space being relevant for the $1S$ exciton.

Despite of all these modifications,
it is important to note that the Hamiltonian is essentially the same as the one 
being applied to the odd exciton system~\cite{100}. The fundamental modifications
presented in this paper are irrelevant for the odd parity system because 
of their $\delta$ function like nature or their specific form affecting only
exciton states with a small radius. 
Hence, we present a consistent theoretical model for the complete exciton
spectrum of $\mathrm{Cu_{2}O}$.

Comparing our results to experimental data, we can prove very good agreement
as regards not only the energies but also the oscillator strengths
since our method of solving the Schr\"odinger equation
allows us also to calculate relative oscillator strengths for 
one-photon and two-photon absorption.
This agreement between theory and experiment
is important not only for the investigation of exciton spectra
in electric or combined electric and magnetic fields.
A correct theoretical description of excitons 
is indispensable if Rydberg 
excitons will be used in the future in quantum information 
technology, or used to attain a deeper understanding 
of quasi-particle interactions in semiconductors~\cite{77,GRE}.
Furthermore, this agreement is a prerequisite for a 
future search for exceptional points in the 
exciton spectrum~\cite{50}.

The paper is organized as follows: 
Having presented the Hamiltonian
of excitons in $\mathrm{Cu_{2}O}$ when considering the complete valence band structure
Sec.~\ref{sec:Hamiltonian}, we discuss all corrections to this
Hamiltonian due to the small radius of the $1S$ exciton in Sec.~\ref{sec:ccc}.
In Sec.~\ref{sec:eosc} we show how to solve the Schr\"odinger equation
using a complete basis and how to calculate relative oscillator strengths 
for one-photon and two-photon absorption.
In Sec.~\ref{sec:results} we discuss the complete yellow and green exciton 
spectrum of $\mathrm{Cu_{2}O}$ paying attention to the exciton 
states with a small principal quantum number
and especially to the green $1S$ exciton state.
Finally, we give a short summary and outlook in Sec.~\ref{sec:Summary-and-outlook}.

\section{Hamiltonian~\label{sec:Hamiltonian}}

In this section we present the Hamiltonian
of excitons in $\mathrm{Cu_{2}O}$,
which accounts for the complete valence band structure
of this semiconductor.
This Hamiltonian describes the exciton states of
odd parity with a principal quantum number $n\geq 3$ very
well~\cite{100,125}. However, for the exciton states of
even parity and for the $2P$ exciton corrections to this
Hamiltonian are needed, which will be described in Sec.~\ref{sec:ccc}.

The lowest $\Gamma_6^+$ conduction band in $\mathrm{Cu_{2}O}$
is almost parabolic in the vicinity of the $\Gamma$ point
and the kinetic energy can be described by the simple
expression
\begin{equation}
H_{\mathrm{e}}\!\left(\boldsymbol{p}_{\mathrm{e}}\right)=\frac{\boldsymbol{p}_{\mathrm{e}}^{2}}{2m_{\mathrm{e}}},\label{eq:He}
\end{equation}
with the effective electron mass $m_{\mathrm{e}}$.
Since $\mathrm{Cu_{2}O}$ has cubic symmetry, we use the
irreducible representations $\Gamma_{i}^{\pm}$
of the cubic group $O_{\mathrm{h}}$ to assign the symmetry of the bands.

Due to interband interactions and nonparabolicities
of the three uppermost valence bands in $\mathrm{Cu_{2}O}$,
the kinetic energy of the hole is given 
by the more complex expression~\cite{80,100,125},
\begin{eqnarray}
H_{\mathrm{h}}\!\left(\boldsymbol{p}_{\mathrm{h}}\right) & = & H_{\mathrm{so}}+\left(1/2\hbar^{2}m_{0}\right)\left\{ \hbar^{2}\left(\gamma_{1}+4\gamma_{2}\right)\boldsymbol{p}_{\mathrm{h}}^{2}\right.\phantom{\frac{1}{1}}\nonumber \\
 & + & 2\left(\eta_{1}+2\eta_{2}\right)\boldsymbol{p}_{\mathrm{h}}^{2}\left(\boldsymbol{I}\cdot\boldsymbol{S}_{\mathrm{h}}\right)\phantom{\frac{1}{1}}\nonumber \\
 & - & 6\gamma_{2}\left(p_{\mathrm{h}1}^{2}\boldsymbol{I}_{1}^{2}+\mathrm{c.p.}\right)-12\eta_{2}\left(p_{\mathrm{h}1}^{2}\boldsymbol{I}_{1}\boldsymbol{S}_{\mathrm{h}1}+\mathrm{c.p.}\right)\phantom{\frac{1}{1}}\nonumber \\
 & - & 12\gamma_{3}\left(\left\{ p_{\mathrm{h}1},p_{\mathrm{h}2}\right\} \left\{ \boldsymbol{I}_{1},\boldsymbol{I}_{2}\right\} +\mathrm{c.p.}\right)\phantom{\frac{1}{1}}\nonumber \\
 & - & \left.12\eta_{3}\left(\left\{ p_{\mathrm{h}1},p_{\mathrm{h}2}\right\} \left(\boldsymbol{I}_{1}\boldsymbol{S}_{\mathrm{h}2}+\boldsymbol{I}_{2}\boldsymbol{S}_{\mathrm{h}1}\right)+\mathrm{c.p.}\right)\right\} \phantom{\frac{1}{1}}\label{eq:Hh}
\end{eqnarray}
with $\left\{ a,b\right\} =\frac{1}{2}\left(ab+ba\right)$, the 
free electron mass $m_{0}$, and c.p.~denoting cyclic permutation. 
The quasi-spin $I=1$
describes the threefold degenerate valence band and
is a convenient abstraction to denote the three 
orbital Bloch functions $xy$, $yz$, and $zx$~\cite{25}.
The parameters $\gamma_{i}$, which are called the first three Luttinger parameters,
and the parameters $\eta_{i}$ describe the behavior and the 
anisotropic effective mass of the hole in the vicinity of the $\Gamma$ point.
The spin-orbit coupling, which enters Eq.~(\ref{eq:Hh}), is given by
\begin{equation}
H_{\mathrm{so}}=\frac{2}{3}\Delta\left(1+\frac{1}{\hbar^{2}}\boldsymbol{I}\cdot\boldsymbol{S}_{\mathrm{h}}\right).\label{eq:soc}
\end{equation}

In a first approximation, the interaction
between the electron and the hole is described
by a screened Coulomb potential
\begin{equation}
V\!\left(\boldsymbol{r}_{e}-\boldsymbol{r}_{h}\right)=-\frac{e^{2}}{4\pi\varepsilon_{0}\varepsilon_{\mathrm{s}1}}\frac{1}{\left|\boldsymbol{r}_{e}-\boldsymbol{r}_{h}\right|}
\end{equation}
with the dielectric constant $\varepsilon_{\mathrm{s}1}=7.5$.
For small relative distances $r=\left|\boldsymbol{r}\right|=\left|\boldsymbol{r}_{e}-\boldsymbol{r}_{h}\right|$
corrections to this potential and to the kinetic energies $H_{\mathrm{e}}\left(\boldsymbol{p}_{\mathrm{e}}\right)$ and $H_{\mathrm{h}}\left(\boldsymbol{p}_{\mathrm{h}}\right)$
are needed, which will be described in Sec.~\ref{sec:ccc}
and which will be denoted here by $V^{\mathrm{CCC}}$.

After introducing relative and center of mass coordinates~\cite{90}
and setting the position and momentum of the 
center of mass to zero, the complete Hamiltonian
of the relative motion finally reads~\cite{17_17,7}
\begin{eqnarray}
H & = & E_{\mathrm{g}}+V\!\left(\boldsymbol{r}\right)+H_{\mathrm{e}}\!\left(\boldsymbol{p}\right)+H_{\mathrm{h}}\!\left(\boldsymbol{p}\right)+V_{\mathrm{CCC}}\label{eq:H}
\end{eqnarray}
with the energy $E_{\mathrm{g}}$ of the band gap.

Note that by setting the total momentum to zero, we neglect polariton effects,
even though in experiments the polaritonic part is always present.
However, when considering the experimental results
of Refs.~\cite{9_1,9,8}, the 
polariton effect on the $1S$ exciton is on the order of 
tens of $\mu\mathrm{eV}$ and, hence, much smaller than the
energy shifts considered here.
Furthermore, in Ref.~\cite{83}
criteria for the experimental observability of polariton 
effects are given. Inserting the material parameters of 
$\mathrm{Cu_{2}O}$ and the experimental linewidths of the 
exciton states observed in Refs.~\cite{GRE,28}, it can be shown that 
polariton effects are not observable for the exciton states of $n\geq 2$.
We will discuss this in greater detail in a future work.

\section{Central-cell corrections~\label{sec:ccc}}

Due to its small radius, the $1S$ exciton in $\mathrm{Cu_{2}O}$ 
is an exciton intermediate
between a Frenkel and a Wannier exciton~\cite{TOE}.
Hence, appropriate corrections
are needed to describe this exciton state
correctly. The corrections, which 
allow for the best possible description of the exciton
problem within the continuum approximation of the solid,
are called central-cell corrections
and have first been treated by Uihlein~\emph{et~al}~\cite{6,7} and
Kavoulakis~\emph{et~al}~\cite{1} for $\mathrm{Cu_{2}O}$.
While Uihlein~\emph{et~al}~\cite{7} accounted for these corrections
only in a simplified way by using a semi-empirical contact potential
$V=-V_0\delta\!\left(\boldsymbol{r}\right)$, the treatment 
of Kavoulakis~\emph{et~al}~\cite{1} did non account for the band structure
and the effect of the central-cell corrections was discussed
only on the $1S$ state and only using perturbation theory.
By considering the complete valence band structure of 
$\mathrm{Cu_{2}O}$ in combination with a non-perturbative 
treatment of the central-cell corrections,
we present a more accurate treatment of the whole
yellow exciton series in $\mathrm{Cu_{2}O}$.
Corrections beyond the frame of the
continuum approximation will not be treated here. 
However, these corrections may
describe remaining small deviations between experimental
and theoretical results.

The central-cell corrections as discussed in Ref.~\cite{1}
comprise three effects, which are
(i) the appearance of terms of higher-order in the momentum
$\boldsymbol{p}$ in the kinetic energies of electron and hole,
(ii) the momentum- and frequency-dependence of the dielectric function $\varepsilon$, 
and (iii) the appearance of an exchange interaction, which depends on the 
momentum of the center of mass.

\subsection{Band structure of Cu$_2$O~\label{sec:Bandstruc}}

Since the radius of the yellow $1S$ exciton is small,
the extension of its wave function in momentum space
is accordingly large. Hence, we have to consider
terms of the fourth power of $\boldsymbol{p}$ in
the kinetic energy of the electron and the hole.
The inclusion of $p^4$ terms in Eqs.~(\ref{eq:He}) and~(\ref{eq:Hh})
leads to an extended and modified Hamiltonian
in the sense of Altarelli, Baldereschi and 
Lipari~\cite{17_17_18,17_17_26,7_11,17_17,17_15}
or Suzuki and Hensel~\cite{44_12}.

The extended Hamiltonian must be compatible with the
symmetry $O_{\mathrm{h}}$ of the crystal and transform
according to the irreducible representation $\Gamma_1^+$.
All the terms of the fourth power of $\boldsymbol{p}$
span a fifteen-dimensional space with the basis functions
\begin{equation}
p_i^4,\quad p_i^3 p_j,\quad p_i^2 p_j^2,\quad p_i p_j p_k^2
\end{equation}
with $i,j,k\in\{1,2,3\}$ and $i\neq j\neq k\neq i$.
Including the quasi spin $I$ and using group theory, one can find six 
linear combinations of $p^4$ terms, which transform according to 
$\Gamma_1^+$~\cite{G3} (see Appendix~\ref{sec:p4}).
Using the results of Appendix~\ref{sec:p4}, we can write the kinetic
energy of the electron and the hole as
\begin{widetext}
\begin{eqnarray}
H_{\mathrm{e}}\!\left(\boldsymbol{p}_{\mathrm{e}}\right) & = & \frac{1}{2\hbar^2 m_{\mathrm{e}}}\left\{\left(\hbar^2+\lambda_{1}a^2\boldsymbol{p}_{\mathrm{e}}^{2}\right)\boldsymbol{p}_{\mathrm{e}}^{2}+\lambda_{2}a^2\left[p_{\mathrm{e}1}^2 p_{\mathrm{e}2}^2+\mathrm{c.p.}\right]\right\}\label{eq:Hecorr}
\end{eqnarray}
and
\begin{eqnarray}
H_{\mathrm{h}}\!\left(\boldsymbol{p}_{\mathrm{h}}\right) & = & H_{\mathrm{so}}+\frac{1}{2\hbar^{4}m_{0}}\left\{ \left(\gamma_{1}+4\gamma_{2}\right)\hbar^{2}\left(\hbar^{2}+\xi_{1}a^{2}\boldsymbol{p}_{\mathrm{h}}^{2}\right)\boldsymbol{p}_{\mathrm{h}}^{2}+\xi_{2}a^{2}\hbar^{2}\left[p_{\mathrm{h}1}^{2}p_{\mathrm{h}2}^{2}+\mathrm{c.p.}\right]\right.\nonumber \\
\nonumber \\
 & - & 6\gamma_{2}\left(\hbar^{2}+\xi_{3}a^{2}\boldsymbol{p}_{\mathrm{h}}^{2}\right)\left[p_{\mathrm{h}1}^{2}\boldsymbol{I}_{1}^{2}+\mathrm{c.p.}\right]-12\gamma_{3}\left(\hbar^{2}+\xi_{4}a^{2}\boldsymbol{p}_{\mathrm{h}}^{2}\right)\left[p_{\mathrm{h}1}p_{\mathrm{h}2}\left\{ \boldsymbol{I}_{1},\boldsymbol{I}_{2}\right\} +\mathrm{c.p.}\right]\nonumber \\
\nonumber \\
 & + & 2\left(\eta_{1}+2\eta_{2}\right)\hbar^{2}\left[\boldsymbol{p}_{\mathrm{h}}^{2}\;\boldsymbol{I}\cdot\boldsymbol{S}_{\mathrm{h}}\right]-12\eta_{2}\hbar^{2}\left[p_{\mathrm{h}1}^{2}\boldsymbol{I}_{1}\boldsymbol{S}_{\mathrm{h}1}+\mathrm{c.p.}\right]-12\eta_{3}\hbar^{2}\left[p_{\mathrm{h}1}p_{\mathrm{h}2}\left(\boldsymbol{I}_{1}\boldsymbol{S}_{\mathrm{h}2}+\boldsymbol{I}_{2}\boldsymbol{S}_{\mathrm{h}1}\right)+\mathrm{c.p.}\right]\nonumber \\
\nonumber \\
 & - & \left. 6\xi_{5}a^{2}\left[\left(p_{\mathrm{h}1}^{4}+6p_{\mathrm{h}2}^{2}p_{\mathrm{h}3}^{2}\right)\boldsymbol{I}_{1}^{2}+\mathrm{c.p.}\right]-12\xi_{6}a^{2}\left[\left(p_{\mathrm{h}1}^{2}+p_{\mathrm{h}2}^{2}-6p_{\mathrm{h}3}^{2}\right)p_{\mathrm{h}1}p_{\mathrm{h}2}\left\{ \boldsymbol{I}_{1},\,\boldsymbol{I}_{2}\right\} +\mathrm{c.p.}\right] \right\}\label{eq:Hhcorr}
\end{eqnarray}
\end{widetext}
with the lattice constant $a$ and the unknown parameters $\lambda_i$ and $\xi_i$.
Note that the values of parameters $\eta_i$ are smaller than the Luttinger parameters $\gamma_i$ (see Table~\ref{tab:1}).
Hence, we expect the terms of the form $p^4 IS_{\mathrm{h}}$ to be negligibly small.

\begin{figure}

\includegraphics[width=1.0\columnwidth]{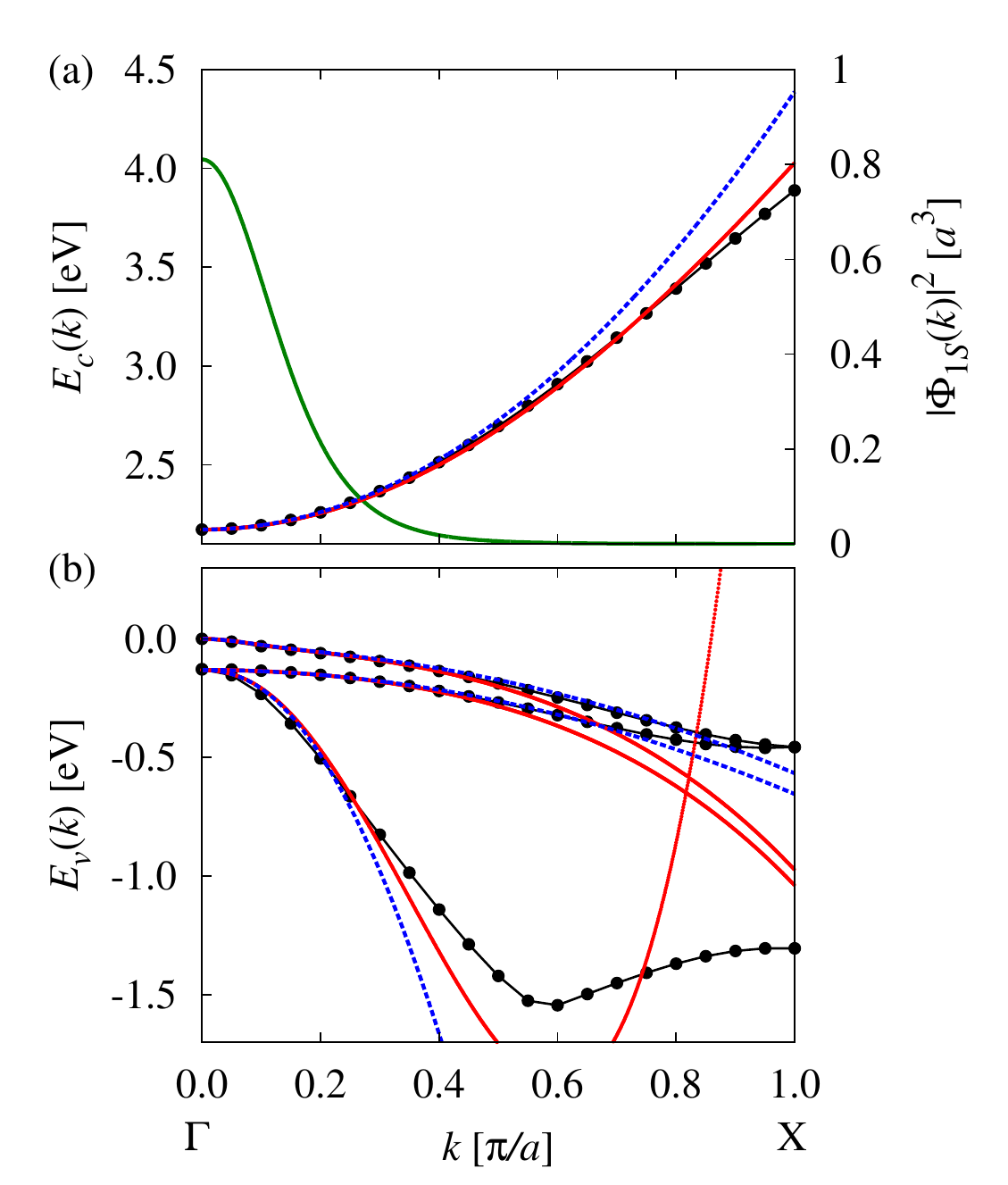}

\caption{Fits to the 
band structure obtained via spin density functional theory calculations~\cite{20} (black linespoints)
for (a) conduction band and (b) valence bands of $\mathrm{Cu_{2}O}$
for the [100] direction 
using the expressions~(\ref{eq:Hecorr}) and~(\ref{eq:Hhcorr}) (red lines).
The green solid line shows the function $\left|\Phi_{1S}\left(\boldsymbol{k}\right)\right|^2$
for $a_{\mathrm{exc}}^{\left(1S\right)}=a$ in units of $a^3$. One can see that the differences between the fit using quartic terms
and the fit of Ref.~\cite{80} (blue dashed lines) neglecting these terms
are small in the range of extension of $\left|\Phi_{1S}\left(\boldsymbol{k}\right)\right|^2$.
Note that $\left|\Phi_{1S}\!\left(\boldsymbol{k}\right)\right|^2$ is not shown in the lower panel for
reasons of clarity.\label{fig:FigGX}}
\end{figure}

\begin{figure}

\includegraphics[width=1.0\columnwidth]{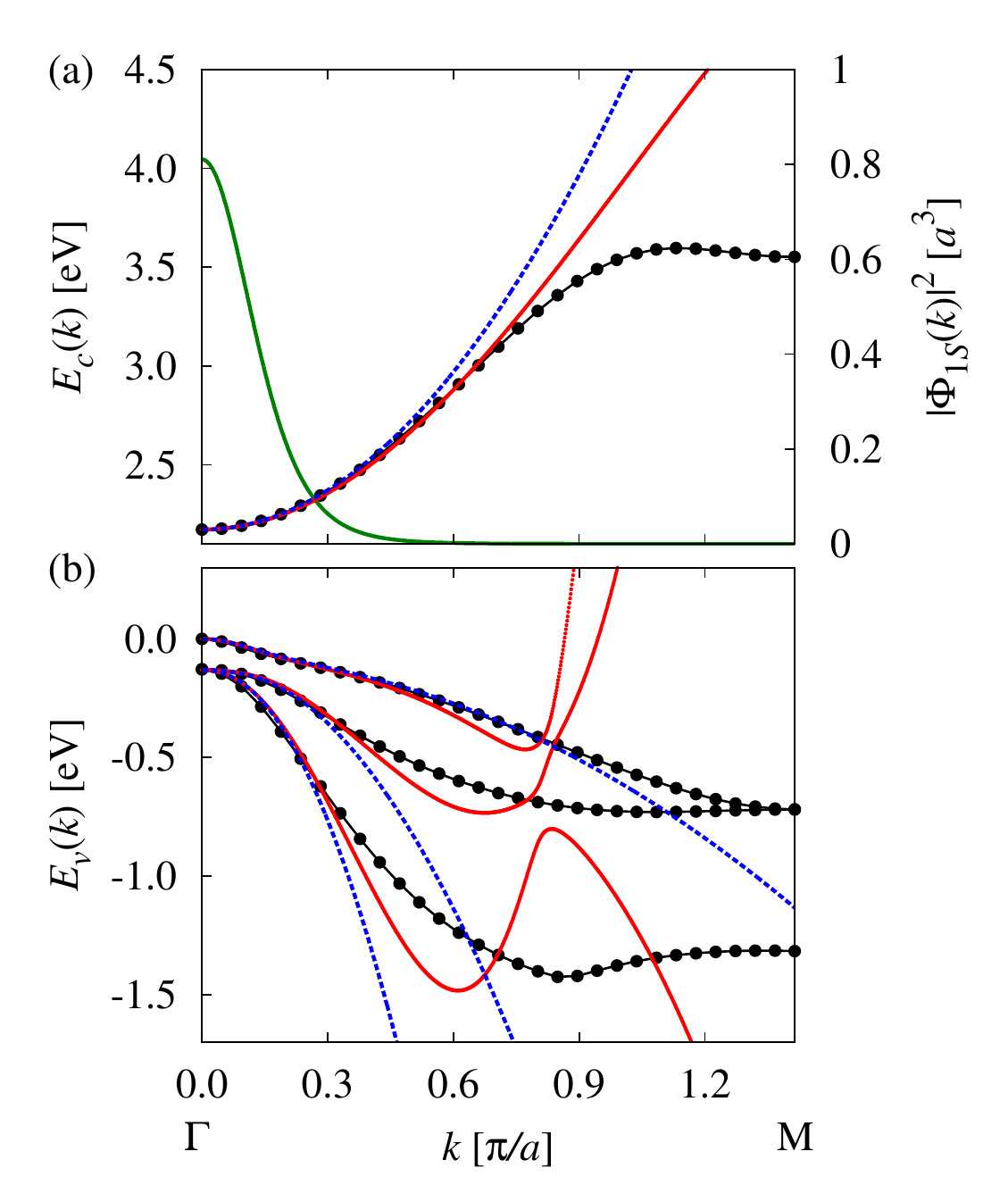}

\caption{Same as Fig.~\ref{fig:FigGX}
for the [110] direction.\label{fig:FigGM}}
\end{figure}

After replacing $H_{\mathrm{e}}\!\left(\boldsymbol{p}_{\mathrm{e}}\right)\rightarrow H_{\mathrm{e}}\!\left(\hbar\boldsymbol{k}\right)$
and $H_{\mathrm{h}}\!\left(\boldsymbol{p}_{\mathrm{h}}\right)\rightarrow -H_{\mathrm{h}}\!\left(\hbar\boldsymbol{k}\right)$, we can determine
the eigenvalues of these Hamiltonians and fit them as in 
Ref.~\cite{80} for $\left|\boldsymbol{k}\right|<\pi/a$
to the band structure of $\mathrm{Cu_{2}O}$ obtained via spin density functional theory calculations~\cite{20}.

To obtain a reliable result, we perform a least-squares fit
with a weighting function. 
Even though the exciton ground state
will show deviations from a pure hydrogen-like $1S$ state, we expect
that the radial probability density can be 
described qualitatively by that function.
Hence, we use the modulus squared of the Fourier transform 
$\Phi_{1S}\!\left(\boldsymbol{k}\right)=\mathcal{F}\left(\Psi_{1S}\right)\left(\boldsymbol{k}\right)$ of the 
hydrogen-like function 
\begin{equation}
\Psi_{1S}\!\left(\boldsymbol{r}\right)=\frac{1}{\sqrt{\pi \left(a_{\mathrm{exc}}^{\left(1S\right)}\right)^3}}e^{-r/a_{\mathrm{exc}}^{\left(1S\right)}}
\end{equation}
as the weighting function for the fit. It reads~\cite{IPEP}
\begin{eqnarray}
\left|\Phi_{1S}\!\left(\boldsymbol{k}\right)\right|^2 & \sim & \left|\frac{1}{\sqrt{(2\pi)^3}}\int\mathrm{d}\boldsymbol{r}\;\Psi_{1S}\!\left(\boldsymbol{r}\right)e^{-i\boldsymbol{k}\boldsymbol{r}}\right|^2\nonumber \\
\nonumber \\
& = & \frac{8 \left(a_{\mathrm{exc}}^{\left(1S\right)}\right)^3}{\pi^2 \left(1+k^2 \left(a_{\mathrm{exc}}^{\left(1S\right)}\right)^2\right)^4}
\end{eqnarray}
with the radius $a_{\mathrm{exc}}^{\left(1S\right)}$ of the $1S$ exciton state.
Although we do not \emph{a priori} know the true
value of $a_{\mathrm{exc}}^{\left(1S\right)}$, the experimental value of the
binding energy of the $1S$ state~\cite{TOE,7} as well as the
calculations of Ref.~\cite{1} indicate that it is
on the order of one or two times
the lattice constant $a=0.427\,\mathrm{nm}$~\cite{JPCS27,P6,DB_45}.
For the fit to the band structure we assume a 
small value of $a_{\mathrm{exc}}^{\left(1S\right)}=a$ 
as a lower limit in the sense of a safe estimate
since then the extension
of the exciton wave function in Fourier space is larger.
In doing so, we will now show that even if the radius
of the $1S$ exciton were smaller or equal to the 
lattice constant $a$, there would not be contributions of the
$p^4$ terms of the band structure.

\begin{figure}[t]

\includegraphics[width=1.0\columnwidth]{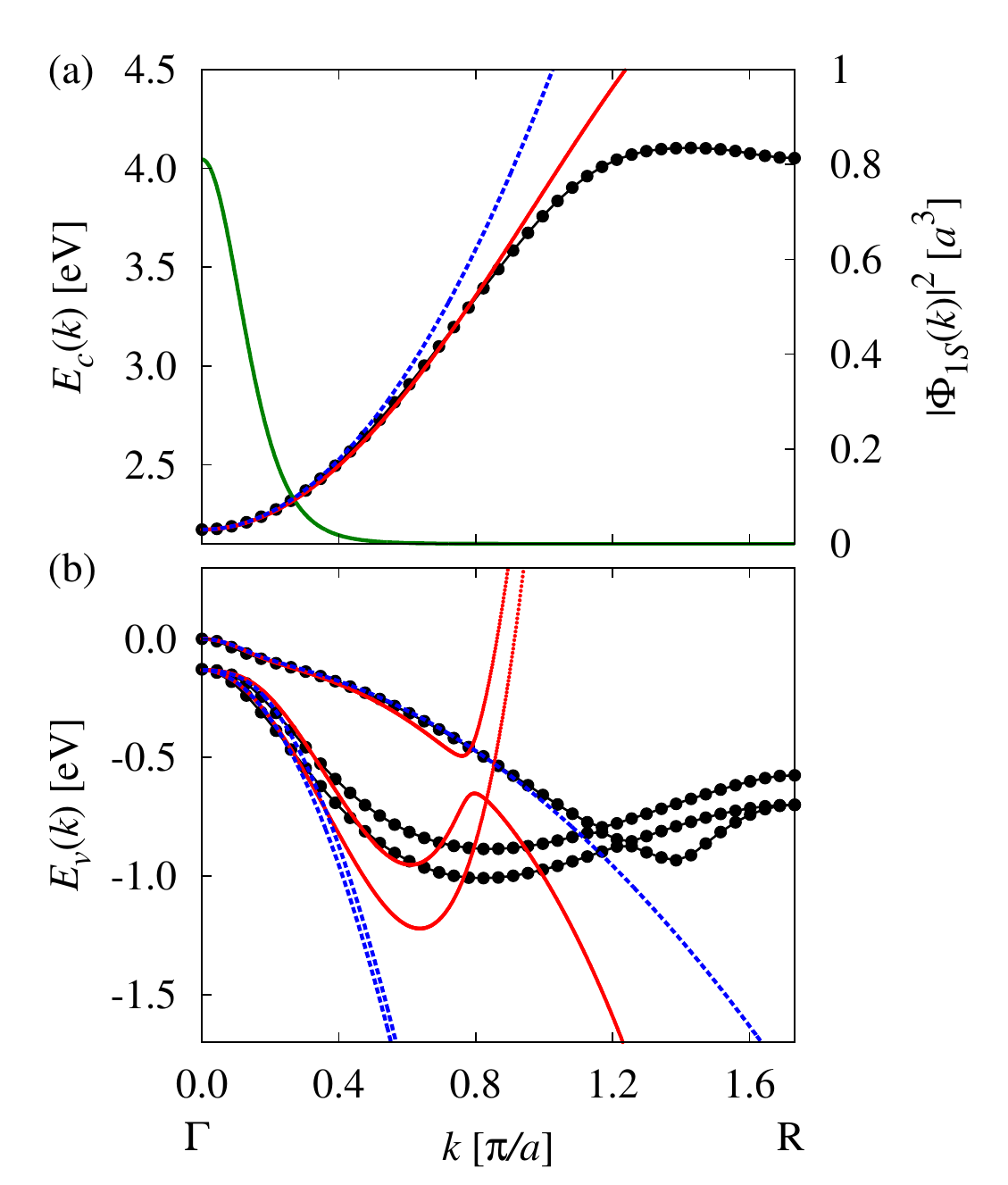}

\caption{Same as Fig.~\ref{fig:FigGX}
for the [111] direction.\label{fig:FigGR}}
\end{figure}

The results of the fit are depicted as red solid lines in Figs.~\ref{fig:FigGX}, \ref{fig:FigGM}, and~\ref{fig:FigGR}.
For a comparison, we also show the fit neglecting the quartic terms
in the momenta (blue dashed lines)~\cite{80}. 
The values of the fit parameters are
\begin{alignat}{4}
\lambda_1 = & -1.109\times 10^{-2} ,\quad          && \lambda_2 = && -2.052\times 10^{-2}, \nonumber \\
\xi_1     = & -1.389\times 10^{-1} ,\quad          && \xi_4     = && -1.518\times 10^{-1}, \nonumber \\
\xi_2     = &  \quad\: 2.353\times 10^{-3} ,\quad  && \xi_5     = &&  \quad\: 9.692\times 10^{-4}, \nonumber \\
\xi_3     = & -1.523\times 10^{-1} ,\quad          && \xi_6     = && -8.385\times 10^{-4}. 
\end{alignat}
As can be seen, e.g., from
Fig.~\ref{fig:FigGM}, the fit including the quartic terms is only slightly better
than the fit with the quadratic terms for small $k$.
A clear difference between the fits can be seen only for large values of $k$
as regards the valence bands:
Since some of the pre-factors of the quartic terms are positive, 
the energy of the valence bands in the fitted model increases
for larger values of $k$.

Considering the minor differences between the fits for small $k$ and
the small extension of the $1S$ exciton function
in $k$ space even for 
$a_{\mathrm{exc}}^{\left(1S\right)}=a$ (see, e.g., Fig.~\ref{fig:FigGX}), the quartic terms
will hardly affect this exciton state and can be neglected.
These arguments still hold if, e.g., 
$a_{\mathrm{exc}}^{\left(1S\right)}=0.2a$ is assumed.

In the work of Ref.~\cite{1} the introduction
of $p^4$ terms seemed necessary to explain the experimentally observed large
mass of the $1S$ exciton. However, the experimental observations are already
well described by quadratic terms in $p$ when considering 
the complete valence band structure~\cite{100}. 
As we already stated
in Ref.~\cite{150}, a simple restriction to the $\Gamma_7^+$ band neglecting the $\Gamma_8^+$
band and considering the nonparabolicity of the $\Gamma_7^+$ band via $p^4$ terms
as has been done in Ref.~\cite{1} does not treat the problem correctly.

\subsection{Dielectric constant~\label{sec:eps}}

In the case of the $1S$ exciton in $\mathrm{Cu_{2}O}$ the relative
motion of the electron and the hole is sufficiently fast that 
phonons cannot follow it and corrections
on the dielectric constant need to be considered.

In general, the electron and the hole
are coupled to longitudinal optical phonons via the 
Fr\"{o}hlich interaction~\cite{AP3,2}
and to longitudinal acoustic phonons via the deformation potential coupling~\cite{M1_7,2}. 
While in the case of optical phonons 
the ions of the solid are displaced in anti-phase 
and thus create a dipole moment in the unit cell of a polar crystal,
the ions are displaced in phase in the case of acoustic phonons
and no dipole moment is created.
Hence, one expects that the interaction
between electron or hole and optical phonons is much larger
than the interaction with acoustic phonons in polar crystals~\cite{HP0,SO}.

If the frequency of the relative motion of electron and hole is
high enough so that the ions of the solid cannot follow it,
the Coulomb interaction between electron and hole
is screened by the high-frequency or background dielectric constant $\varepsilon_{\mathrm{b}}$~\cite{TOE,SOK1_82L1}.
This dielectric constant describes the electronic polarization,
which can follow the motion of electron and hole very quickly~\cite{PAE}.

For lower frequencies of the relative motion the
contribution of the phonons to the screening becomes important
and the dielectric function $\varepsilon$ becomes frequency
dependent. 
In many semiconductors the frequency of the relative motion
in exciton states with a principal quantum number of $n\geq 2$
is so small that the low-frequency or static dielectric constant $\varepsilon_{\mathrm{s}}$
can be used~\cite{SO}, which involves the electronic polarization
\emph{and} the displacement of the ions~\cite{PAE}.
Note that we use the notation $\varepsilon_{\mathrm{b}}$, $\varepsilon_{\mathrm{s}}$
instead of $\varepsilon_{\infty}$, $\varepsilon_{0}$ to avoid
the risk of confusion with the electric permittivity $\varepsilon_{0}$~\cite{PAE,SO}.

The transition from $-e^2/4\pi\varepsilon_{0}\varepsilon_{\mathrm{s}}r$
to $-e^2/4\pi\varepsilon_{0}\varepsilon_{\mathrm{b}}r$, which takes place when the
frequency of the electron or the hole is of the
same size as the frequency of the phonon~\cite{PAE},
had been investigated in detail by
Haken in Refs.~\cite{HP4_21,HP8,HP9,HP0,HP1,PAE}.
He considered at first the interaction between electron or hole
and the phonons and then constructed the exciton from the 
resulting particles with polarisation clouds, i.e., the polarons.
The change of the Coulomb interaction between both particles
was then explained in terms of an exchange of phonons, i.e., of 
virtual quanta of the polarization field~\cite{PAE}.

The final result for the interaction in the transition region
between $-e^2/4\pi\varepsilon\varepsilon_{\mathrm{s}}r$
and $-e^2/4\pi\varepsilon\varepsilon_{\mathrm{b}}r$ was the
so-called Haken potential~\cite{HP8,HP9,HP1,TOE,SO,IPEP},
\begin{eqnarray}
V\!\left(r\right) & = & -\frac{e^2}{4\pi\varepsilon_0 r}\left[\frac{1}{\varepsilon_{\mathrm{s}}}+\frac{1}{2\varepsilon^*}\left(e^{-r/\rho_{\mathrm{h}}}+e^{-r/\rho_{\mathrm{e}}}\right)\right].\label{eq:Haken}
\end{eqnarray}
Here $\rho_{\mathrm{e}}$ and $\rho_{\mathrm{h}}$ denote the polaron
radii
\begin{equation}
\rho_{\mathrm{e/h}}=\sqrt{\frac{\hbar}{2m_{\mathrm{e/h}}^*\omega_{\mathrm{LO}}}}
\end{equation}
with the frequency $\omega_{\mathrm{LO}}$ of the optical phonon
and $\varepsilon^*$ is given by
\begin{equation}
\frac{1}{\varepsilon^*}=\frac{1}{\varepsilon_{\mathrm{b}}}-\frac{1}{\varepsilon_{\mathrm{s}}}.
\end{equation}
Note that in the result of Haken~\cite{HP0,HP4} the polaron masses $m_i^*$ instead of the bare
electron and hole masses have to be used in the polaron radii
and the kinetic energies. 
Furthermore, the lattice relaxation due to the interaction of excitons
and phonons decreases the band gap energy for electrons and holes.
However, since the value of $E_{\mathrm{g}}$ for $\mathrm{Cu_{2}O}$ 
has been determined in Ref.~\cite{GRE} from the experimental exciton spectrum,
the polaron effect is already accounted for in the band gap energy~\cite{SO}.

Note that the above results were derived in the simple band model
and by assuming only one optical phonon branch contributing to the
Fr\"ohlich interaction. To the best of our knowledge there is no model
accounting for more than one optical phonon branch~\cite{HP4,1,PAE}, which
complicates the correct treatment of $\mathrm{Cu_{2}O}$,
where two LO phonons contribute to the Fr\"ohlich interaction.
Even though there are theories for polarons in the degenerate
band case~\cite{22,E3,HP7}, we will use only the leading,
spherically symmetric terms, in which only the isotropic
effective mass of the hole or only the Luttinger parameter $\gamma_1$
enters. Of course, there are further terms of cubic symmetry, which
also depend on the other Luttinger parameters. However, since already
$\gamma_1$ is at least by a factor of $2$ larger 
than the other Luttinger parameters,
we expect the further terms in the Haken potentials to be smaller
than the leading term used here.
Since the effect of the Haken potential on the
exciton spectrum is not crucial, as will be seen from Fig.~6,
the neglection of further terms in the polaron potentials
will then be \emph{a~posteriori} justified.

Furthermore, the Haken potential~(\ref{eq:Haken}) cannot describe the non-Coulombic
electron-hole interaction for very small values of $r$, which is due
to the finite size of electron and hole~\cite{TOE}. The conditions of validity of the
potential~(\ref{eq:Haken}) have, e.g., been discussed by Haken in Ref.~\cite{PAE}.

When treating the Haken potential numerically for different polar crystals,
the experimental and theoretical binding energies of the exciton states
sometimes do not agree, for which reason corrections, sometimes phenomenologically,
to the Haken potential have been introduced~\cite{HP5,HP2,HP3,HP6} leading to 
clearly better results.
One of these refined formulas is the potential proposed by Pollmann and B\"uttner~\cite{HP6,HP4}
\begin{eqnarray}
V\!\left(r\right) & = & -\frac{e^2}{4\pi\varepsilon_0 r}\nonumber \\
\nonumber \\
 & \times & \left[\frac{1}{\varepsilon_{\mathrm{s}}}+\frac{1}{\varepsilon^*}\left(\frac{m_{\mathrm{h}}}{\Delta m}e^{-r/\rho_{\mathrm{h}}}-\frac{m_{\mathrm{e}}}{\Delta m}e^{-r/\rho_{\mathrm{e}}}\right)\right],\label{eq:PB}
\end{eqnarray}
in which the bare electron and hole masses have to be used and 
where $\Delta m$ is given by $\Delta m=m_{\mathrm{h}}-m_{\mathrm{e}}$.
Hence, we take the statements given above as a reason to propose the following
phenomenological potentials for $\mathrm{Cu_{2}O}$, which are motivated by the
formula of Haken and by the formula of Pollmann and B\"uttner:
\begin{widetext}
\begin{subequations}
\begin{eqnarray}
V^{\mathrm{H}}\!\left(r\right)=-\frac{e^{2}}{4\pi\varepsilon_{0}r}\left[\frac{1}{\varepsilon_{\mathrm{s}1}}+\frac{1}{2\varepsilon_{1}^{*}}\left(e^{-r/\rho_{\mathrm{h}1}}+e^{-r/\rho_{\mathrm{e}1}}\right)+\frac{1}{2\varepsilon_{2}^{*}}\left(e^{-r/\rho_{\mathrm{h}2}}+e^{-r/\rho_{\mathrm{e}2}}\right)\right]\label{eq:Haken2}
\end{eqnarray}
and
\begin{eqnarray}
V^{\mathrm{PB}}\!\left(r\right) & = & -\frac{e^{2}}{4\pi\varepsilon_{0}r}\left[\frac{1}{\varepsilon_{\mathrm{s}1}}+\frac{1}{\varepsilon_{1}^{*}}\left(\frac{m_{0}}{m_{0}-m_{\mathrm{e}}\gamma_{1}}e^{-r/\rho_{\mathrm{h}1}}-\frac{m_{\mathrm{e}}\gamma_{1}}{m_{0}-m_{\mathrm{e}}\gamma_{1}}e^{-r/\rho_{\mathrm{e}1}}\right)\right.\nonumber \\		
& &	\;\,\quad\qquad\qquad+\left.\frac{1}{\varepsilon_{2}^{*}}\left(\frac{m_{0}}{m_{0}-m_{\mathrm{e}}\gamma_{1}}e^{-r/\rho_{\mathrm{h}2}}-\frac{m_{\mathrm{e}}\gamma_{1}}{m_{0}-m_{\mathrm{e}}\gamma_{1}}e^{-r/\rho_{\mathrm{e}2}}\right)\right].\label{eq:PB2}
\end{eqnarray}
\end{subequations}
\end{widetext}
Here we use
\begin{equation}
\frac{1}{\varepsilon_{i}^{*}}=\frac{1}{\varepsilon_{\mathrm{b}i}}-\frac{1}{\varepsilon_{\mathrm{s}i}}
\end{equation}
and
\begin{equation}
\rho_{\mathrm{e}i}=\sqrt{\frac{\hbar}{2m_{\mathrm{e}}\omega_{\mathrm{LOi}}}},\qquad\rho_{\mathrm{h}i}=\sqrt{\frac{\hbar\gamma_1}{2m_0\omega_{\mathrm{LOi}}}},
\end{equation}
where the energies of the phonons and the values of the dielectric constants
are given by~\cite{1}
\begin{equation}
\hbar\omega_{\mathrm{LO1}}=18.7\,\mathrm{meV},\qquad\hbar\omega_{\mathrm{LO\,2}}=87\,\mathrm{meV}
\end{equation}
and
\begin{equation}
{\varepsilon_{\mathrm{s}1}}=7.5,\qquad{\varepsilon_{\mathrm{b}1}}={\varepsilon_{\mathrm{s}2}}=7.11,\qquad{\varepsilon_{\mathrm{b}2}}=6.46.
\end{equation}
As has been done in Ref.~\cite{HP4} for perovskite $\mathrm{CH_{3}NH_{3}PbI_{3}}$,
we use $V_{\mathrm{H}}$ or $V_{\mathrm{PB}}$ in the Schr\"odinger equation without an
additional fit parameter and find out which of these potentials describes the
exciton spectrum of $\mathrm{Cu_{2}O}$ best.
Since for the polaron radii $\rho_{\mathrm{e}}$ and $\rho_{\mathrm{h}}$ 
$1.6a\leq \rho \leq 4.4a$ holds, we expect the Haken or the
Pollmann-B\"uttner potential to have a significant influence on the
exciton states with $n\leq 2$.

As the Fr\"ohlich coupling constant is small in $\mathrm{Cu_{2}O}$, i.e., 
it is $\alpha^{\mathrm{F}}\lesssim 0.2$ for the two optical phonons
and both the electron and the hole~\cite{20},
the bare electron and hole masses differ from the polaron masses by at most~3\%.
Hence, we can calculate with the bare masses when using 
$V_{\mathrm{H}}$.

Besides the frequency dependence of the dielectric function
also its momentum dependence becomes important if
the exciton radius is on the order of the lattice constant.
This momentum dependence of the dielectric function arises from the
electronic polarization~\cite{1_23,1}.

When treating the excitons of $\mathrm{Cu_{2}O}$ in momentum space, the wave functions of the
$n\geq 2$ states are localized about $k=0$ so that for these states
the $k$ dependence of $\varepsilon$ is not important.
However, for the $1S$ state $a_{1S}\approx a$ holds and 
thus this state is screened by $\varepsilon$
at higher momenta $k$~\cite{1}.
Considering the Coulomb interaction for the $1S$ exciton in $k$ space,
\begin{equation}
V(k,\,\omega)=-\frac{1}{\sqrt{(2\pi)^3}}\,\frac{e^2}{\varepsilon_0\varepsilon(k,\,\omega) k^2},\label{eq:CoulF}
\end{equation}
Kavoulakis~\emph{et~al}~\cite{1} derived a correction term by assuming
\begin{eqnarray}
\frac{1}{\varepsilon(k,\omega)} & \approx & \frac{1}{\varepsilon_{\mathrm{b}}-d(ka)^2}\approx \frac{1}{\varepsilon_{\mathrm{b}}}+\frac{d(ka)^2}{\varepsilon_{\mathrm{b}}^2}\label{eq:epsexp}
\end{eqnarray}
valid for $\,E_{\mathrm{g}}/\hbar\gg \omega\gg \omega_{\mathrm{LO}}$ with a small unknown constant $d$.
Inserting Eq.~(\ref{eq:epsexp}) in Eq.~(\ref{eq:CoulF})
and Fourier transforming the second expression, one obtains
the following correction term to the Coulomb interaction:
\begin{equation}
V_{d}=-da^2 \frac{e^2}{\varepsilon_0\varepsilon_{\mathrm{b}}^2}V_{\mathrm{uc}}\,\delta\!\left(\boldsymbol{r}\right)=-V_0 V_{\mathrm{uc}}\delta\!\left(\boldsymbol{r}\right).\label{eq:deltaV0}
\end{equation}
Following the calculation of Ref.~\cite{1_23} on the dielectric function and using
the lowest $\Gamma_8^-$ conduction band and the highest $\Gamma_7^+$
valence band, Kavoulakis~\emph{et~al}~\cite{1} estimated the value of $d$ to $d\approx 0.18$~\cite{1}.

Note that in general a Kronecker delta would appear in Eq.~(\ref{eq:deltaV0})~\cite{TOE}.
However, as we treat the exciton problem in the continuum approximation,
this Kronecker delta is replaced by the delta function times 
the volume $V_{\mathrm{uc}}=a^3$ of one unit cell. 
Thus, the parameter $V_0$ has the unit of an energy.

We have already stated above that the Haken potential
cannot describe the electron-hole interaction correctly for very small $r$.
Therefore, we now assume that the potential~(\ref{eq:deltaV0})
is not only due to the momentum dependence of the dielectric function
but that it also accounts for deviations from the Haken potential at small $r$.
Hence, we will treat $V_0$ as an unknown fit parameter in the following.


\subsection{Exchange interaction~\label{sec:exchange}}

In the Wannier equation or Hamiltonian of excitons
the exchange interaction is generally not included but
regarded as a correction to the
hydrogen-like solution~\cite{TOE}.
Recently, we have presented a comprehensive discussion
of the exchange interaction in $\mathrm{Cu_{2}O}$~\cite{150}.
We could show, in accordance with Ref.~\cite{1}, that corrections to the exchange interaction
due to a finite momentum $\hbar\boldsymbol{K}$ of the center of mass of the exciton
are negligibly small.
Hence, only the $K$ independent part of the exchange interaction~\cite{7,E2,E3,150}
\begin{equation}
H_{\mathrm{exch}}=J_{0}\left(\frac{1}{4}-\frac{1}{\hbar^{2}}\boldsymbol{S}_{\mathrm{e}}\cdot\boldsymbol{S}_{\mathrm{h}}\right) V_{\mathrm{uc}}\delta\!\left(\boldsymbol{r}\right)\label{eq:Hexch}
\end{equation}
needs to be considered.
Within the simple hydrogen-like model the exchange interaction
would only affect the $nS$ exciton states as these 
states have a nonvanishing probability density at $r=0$.
However, when considering the complete valence band structure,
the exciton states with even or with odd values of $L$ are coupled,
and thus the exchange interaction will affect the whole even exciton series.

It is well known from experiments that the
splitting between the yellow $1S$ ortho and the yellow $1S$ para exciton
amounts to about $12\,\mathrm{meV}$~\cite{1_6a,E2_16,1_6c}.
Hence, we have to choose the value of $\tilde{J}_{0}$ 
such that this splitting is reflected in the 
theoretical spectrum.

\subsection{Summary~\label{sec:sumccc}}

Following the explanations given in Secs.~\ref{sec:eps} and~\ref{sec:exchange}, the term $V_{\mathrm{CCC}}$ in the Hamiltonian of Eq.~(\ref{eq:H}) takes one of the following forms:
\begin{widetext}
\begin{subequations}
\begin{eqnarray}
V_{\mathrm{CCC}}^{\mathrm{H}}\!\left(\boldsymbol{r}\right) & = & -\frac{e^{2}}{4\pi\varepsilon_{0}r}\left[\frac{1}{2\varepsilon_{1}^{*}}\left(e^{-r/\rho_{\mathrm{h}1}}+e^{-r/\rho_{\mathrm{e}1}}\right)+\frac{1}{2\varepsilon_{2}^{*}}\left(e^{-r/\rho_{\mathrm{h}2}}+e^{-r/\rho_{\mathrm{e}2}}\right)\right]\nonumber\\
& & +\left[-V_0 +J_{0}\left(\frac{1}{4}-\frac{1}{\hbar^{2}}\boldsymbol{S}_{\mathrm{e}}\cdot\boldsymbol{S}_{\mathrm{h}}\right)\right]V_{\mathrm{uc}}\delta\!\left(\boldsymbol{r}\right),\label{eq:HCCC_H}\\
\nonumber\\
\nonumber\\
V_{\mathrm{CCC}}^{\mathrm{PB}}\!\left(\boldsymbol{r}\right) & = & -\frac{e^{2}}{4\pi\varepsilon_{0}r}\left[\frac{1}{\varepsilon_{1}^{*}}\left(\frac{m_{0}}{m_{0}-m_{\mathrm{e}}\gamma_{1}}e^{-r/\rho_{\mathrm{h}1}}-\frac{m_{\mathrm{e}}\gamma_{1}}{m_{0}-m_{\mathrm{e}}\gamma_{1}}e^{-r/\rho_{\mathrm{e}1}}\right)\right.\nonumber \\		
& &	\;\,\quad\qquad\qquad+\left.\frac{1}{\varepsilon_{2}^{*}}\left(\frac{m_{0}}{m_{0}-m_{\mathrm{e}}\gamma_{1}}e^{-r/\rho_{\mathrm{h}2}}-\frac{m_{\mathrm{e}}\gamma_{1}}{m_{0}-m_{\mathrm{e}}\gamma_{1}}e^{-r/\rho_{\mathrm{e}2}}\right)\right]\nonumber\\
& & +\left[-V_0 +J_{0}\left(\frac{1}{4}-\frac{1}{\hbar^{2}}\boldsymbol{S}_{\mathrm{e}}\cdot\boldsymbol{S}_{\mathrm{h}}\right)\right]V_{\mathrm{uc}}\delta\!\left(\boldsymbol{r}\right),\label{eq:HCCC_PB}
\end{eqnarray}
\label{eq:HCCC}
\end{subequations}
\end{widetext}
[cf. Eqs.~(\ref{eq:Haken2}), (\ref{eq:PB2}), (\ref{eq:deltaV0}), and~(\ref{eq:Hexch})].
Note that while the operators with $\delta\left(\boldsymbol{r}\right)$ affect only
the exciton series with even values of $L$, the Haken or 
Pollmann and B\"uttner potential affect all exciton states~\cite{PAE}.
A comparison of our results with the experimental values
of Refs.~\cite{GRE,7,28,DB_49,HO} will allow us, in Sec.~\ref{sec:results},
to determine the size of the unknown parameters $V_0$ and $J_0$.  
 

\section{Eigenvalues and oscillator strengths~\label{sec:eosc}}

In this section we describe how the Schr\"odinger equation
corresponding to the Hamiltonian~(\ref{eq:H})
is solved in a complete basis. Furthermore, we discuss how to calculate
oscillator strengths for two-photon absorption.
An appropriate basis to solve the Schr\"odinger equation
has been presented in detail in Ref.~\cite{100}. Hence, we
recapitulate only the most important points.

As regards the angular momentum part of the basis, we have to
consider that the different operators in the Hamiltonian couple the
quasi spin $I$, the hole spin $S_{\mathrm{h}}$, and the angular momentum $L$
of the exciton. Hence, we introduce the
effective hole spin $J=I+S_{\mathrm{h}}$, the angular momentum $F=L+J$,
and the total angular momentum $F_{t}=F+S_{\mathrm{e}}$.
For the radial
part of the exciton wave function we use the Coulomb-Sturmian functions~\cite{S1}
\begin{equation}
U_{NL}\!\left(r\right)=N_{NL}^{(\alpha)}\left(2\rho\right)^{L}e^{-\rho}L_{N}^{2L+1}\left(2\rho\right)\label{eq:U}
\end{equation}
with $\rho=r/\alpha$, an arbitrary convergence or scaling parameter $\alpha$,
the associated Laguerre polynomials $L_{n}^{m}\left(x\right)$, and 
a normalization factor 
\begin{equation}
N_{NL}^{(\alpha)}=\frac{2}{\sqrt{\alpha^3}}\left[\frac{N!}{\left(N+L+1\right)\left(N+2L+1\right)!}\right]^{\frac{1}{2}}.\label{eq:normaliz}
\end{equation}
The radial quantum number
$N$ is related to the principal quantum number $n$ via $n=N+L+1$.
Finally, we use the following ansatz for the exciton wave function
\begin{subequations}
\begin{eqnarray}
\left|\Psi\right\rangle  & = & \sum_{NLJFF_{t}M_{F_{t}}}c_{NLJFF_{t}M_{F_{t}}}\left|\Pi\right\rangle,\\
\nonumber \\
\left|\Pi\right\rangle  & = & \left|N,\, L;\,\left(I,\, S_{\mathrm{h}}\right)\, J;\, F,\, S_{\mathrm{e}};\, F_{t},\, M_{F_{t}}\right\rangle\label{eq:basis}
\end{eqnarray}\label{eq:ansatz}%
\end{subequations}
with real coefficients $c$. The parenthesis and semicolons in
Eq.~(\ref{eq:basis}) are meant to illustrate the coupling scheme of the
spins and the angular momenta.
Since the $z$ axis is a fourfold axis,
it is sufficient to use only $M_{F_t}$ quantum numbers
which differ by $\pm 4$ in Eq.~(\ref{eq:ansatz}).

We now express the Hamiltonian~(\ref{eq:H}) in terms of irreducible tensors~\cite{ED,7_11,44}.
Inserting the ansatz~(\ref{eq:ansatz}) in the Schr\"odinger
equation $H\Psi=E\Psi$ and multiplying from the left with
another basis state $\left\langle \Pi'\right|$,
we obtain a matrix representation 
of the Schr\"odinger equation of the form
\begin{equation}
\boldsymbol{D}\boldsymbol{c}=E\boldsymbol{M}\boldsymbol{c}.\label{eq:gev}
\end{equation}
The vector $\boldsymbol{c}$ contains the coefficients of the ansatz~(\ref{eq:ansatz}).
All matrix elements, which enter the symmetric matrices $\boldsymbol{D}$ and
$\boldsymbol{M}$ and which have not been treated in Ref.~\cite{100},
are given in Appendix~\ref{sub:Matrix-elements}.
The generalized eigenvalue problem~(\ref{eq:gev})
is finally solved using an appropriate LAPACK routine~\cite{Lapack}.
The material parameters used in our calculation are listed in Table~\ref{tab:1}.

Since the basis cannot be infinitely large, the values of the quantum numbers
are chosen in the following way: For each value of $n=N+L+1$ we use
\begin{eqnarray}
L & = & 0,\,\ldots,\, n-1,\nonumber \\
J & = & 1/2,\,3/2,\nonumber \\
F & = & \left|L-J\right|,\,\ldots,\,\min\left(L+J,\, F_{\mathrm{max}}\right),\\
F_{t} & = & F-1/2,\, F+1/2,\nonumber \\
M_{F_{t}} & = & -F_{t},\,\ldots,\, F_{t}.\nonumber 
\end{eqnarray}
The value $F_{\mathrm{max}}$ and the maximum value of $n$ are chosen
appropriately large so that the eigenvalues converge. Additionally,
we can use the scaling parameter $\alpha$ to enhance convergence.
However, it should be noted that the value of $\alpha$ does not influence the theoretical
results for the exciton energies in any way, i.e., the converged
results do not depend on the value of $\alpha$.

\begin{table}

\protect\caption{Material parameters used in the calculations.
Instead of the band gap energy $E_{\mathrm{g}}=2.17208\,\mathrm{eV}$ of Ref.~\cite{GRE}
a slightly smaller value is used to obtain a better agreement with
experimental values in Sec.~\ref{sec:results}.\label{tab:1}}

\begin{centering}
\begin{tabular}{lll}
\hline 
band gap energy & $E_{\mathrm{g}}=2.17202\,\mathrm{eV}$ & \tabularnewline
electron mass & $m_{\mathrm{e}}=0.99\, m_{0}$ & \cite{M2}\tabularnewline
spin-orbit coupling & $\Delta=0.131\,\mathrm{eV}$ & \cite{80}\tabularnewline
valence band parameters & $\gamma_{1}=1.76$ & \cite{80,100}\tabularnewline
 & $\gamma_2=0.7532$ & \cite{80,100}\tabularnewline
 & $\gamma_3=-0.3668$ & \cite{80,100}\tabularnewline
 & $\eta_1=-0.020$ & \cite{80,100}\tabularnewline 
 & $\eta_2=-0.0037$ & \cite{80,100}\tabularnewline
 & $\eta_3=-0.0337$ & \cite{80,100}\tabularnewline
lattice constant & $a=0.42696\,\mathrm{nm}$ & \cite{20_24}\tabularnewline
dielectric constants & $\varepsilon_{\mathrm{s}1}=7.5$ & \cite{SOK1_82L1}\tabularnewline
 & $\varepsilon_{\mathrm{b}1}=\varepsilon_{\mathrm{s}2}=7.11$ & \cite{SOK1_82L1}\tabularnewline
 & $\varepsilon_{\mathrm{b}2}=6.46$ & \cite{SOK1_82L1}\tabularnewline
energy of $\Gamma_{4}^{-}$-LO phonons & $\hbar\omega_{\mathrm{LO1}}=18.7\,\mathrm{meV}$ & \cite{1}\tabularnewline
 & $\hbar\omega_{\mathrm{LO2}}=87\,\mathrm{meV}$ & \cite{1}\tabularnewline
\hline 
\end{tabular}
\par\end{centering}

\end{table}


Note that the presence of the delta functions in Eq.~(\ref{eq:HCCC})
makes the whole problem more complicated than in Ref.~\cite{100} since not only the eigenvalues
but also the wave functions at $r=0$ have to converge. However,
for a specific value of $\alpha$ it is not possible to obtain convergence for
all exciton states of interest. 
Therefore, we solve the Schr\"odinger
equation initially without the $\delta\!\left(\boldsymbol{r}\right)$ dependent terms.
We then select the converged eigenvectors and with these we set up a
second generalized eigenvalue problem now including
the $\delta\!\left(\boldsymbol{r}\right)$ dependent terms. 
This problem is again solved using an appropriate LAPACK routine~\cite{Lapack}
and provides the correct converged eigenvalues of the complete Hamiltonian~(\ref{eq:H}).

Having solved the eigenvalue problem, we can use the eigenvectors
to determine relative oscillator strengths.
The determination of relative oscillator strengths in one-photon 
absorption has been presented in detail in Refs.~\cite{100,125}.
While in one photon absorption excitons of symmetry $\Gamma_4^-$
are dipole-allowed~\cite{100},
the selection rules for two-photon absorption~\cite{6_15a,6_15b,6_15c} are different
and excitons of symmetry $\Gamma_5^+$
can be optically excited.

When considering one-photon absorption one generally treats the
operator $\boldsymbol{A}\boldsymbol{p}$ with the vector potential $\boldsymbol{A}$
of the radiation field in first order perturbation theory.
The dipole operator then transforms according to the irreducible representation
$D^1$ of the full rotation group. 
In two-photon absorption one needs the operator $\boldsymbol{A}\boldsymbol{p}$
twice and thus the product
$D^1\otimes D^1=D^0 \oplus D^1 \oplus D^2$ has to be considered~\cite{G3}. 
In $\mathrm{Cu_{2}O}$ the reduction of these irreducible representations
by the cubic group $O_{\mathrm{h}}$ has to be considered and one obtains
\begin{equation}
\Gamma_4^-\otimes\Gamma_4^-=\Gamma_1^+\oplus\Gamma_4^+\oplus\left(\Gamma_3^+\oplus\Gamma_5^+\right).
\end{equation}
In two-photon absorption the spin $S=S_{\mathrm{e}}+S_{\mathrm{h}}=0$ remains
unchanged and the exciton state must have an $L=0$ component.
Hence, the correct expression for the relative oscillator strength
is given by 
\begin{equation}
f_{\mathrm{rel}}\sim\left|\lim_{r\rightarrow0}\,_{T}\left\langle 1,\,M'_{F_t}\middle|\Psi\left(\boldsymbol{r}\right)\right\rangle\right|^2,\label{eq:frelT}
\end{equation}
with the wave function $\left|\Psi\right\rangle$ of Eq.~(\ref{eq:ansatz}) and the state
$\left|F'_t,\,M'_{F_t}\right\rangle_T$, which is a short notation for
\begin{eqnarray}
& & \left|\left(S_{\mathrm{e}},\,S_{\mathrm{h}}\right)\,S,\,I;\,I+S,\,L;\,F'_t,\,M'_{F_t}\right\rangle\nonumber\\
& = & \left|\left(1/2,\,1/2\right)\,0,\,1;\,1,\,0;\,F'_t,\,M'_{F_t}\right\rangle.\label{eq:stateT}
\end{eqnarray}
Note that the coupling scheme of the spins and angular momenta in Eq.~(\ref{eq:stateT}) 
given by 
\begin{equation}
S_{\mathrm{e}}+S_{\mathrm{h}}=S\quad\rightarrow\quad(I+S)+L=F'_t
\end{equation}
is different from the one of Eq.~(\ref{eq:basis}) due to the 
requirement that $S$ must be a good quantum number.

It can be shown that the state $\left|1,\,M'_{F_t}\right\rangle_T$ transforms according
to the irreducible representation $\Gamma_5^+$ of $O_{\mathrm{h}}$~\cite{G3}, for which
reason only exciton states of this symmetry can be excited in two-photon absorption.
By choosing particular directions of the polarization of the light, e.g.,
by choosing one photon being polarized in $x$ direction and one photon being polarized
in $y$ direction, only one component of the $\Gamma_5^+$ exciton states, the $xy$ component, can be excited optically.
We consider this case in the following and hence use $M'_{F_t}=0$ in Eq.~(\ref{eq:frelT}).
Finally, we wish to note that the exciton states
of symmetry $\Gamma_5^+$ can weakly be observed in one-photon absorption
in quadrupole approximation~\cite{7}.

\section{Results and discussion~\label{sec:results}}

In this section we determine the values of the parameters
$J_0$ and $V_0$ and discuss the complete exciton spectrum of $\mathrm{Cu_{2}O}$.

The parameter $J_0$ describes the strength of the exchange interaction.
It is well known that the exchange interaction mainly affects
the $1S$ exciton and that the splitting between the ortho
and the para exciton state amounts to $11.8\,\mathrm{meV}$~\cite{1_6a,E2_16,1_6c,7,DB_49}.
By choosing 
\begin{equation}
J_0=0.792\pm 0.068\,\mathrm{eV}\label{eq:J0}
\end{equation}
we obtain the correct
value of this splitting
irrespective of whether using the Haken or
the Pollman-B\"uttner potential [cf. Eq.~(\ref{eq:HCCC})].

\begin{figure}

\includegraphics[width=1.0\columnwidth]{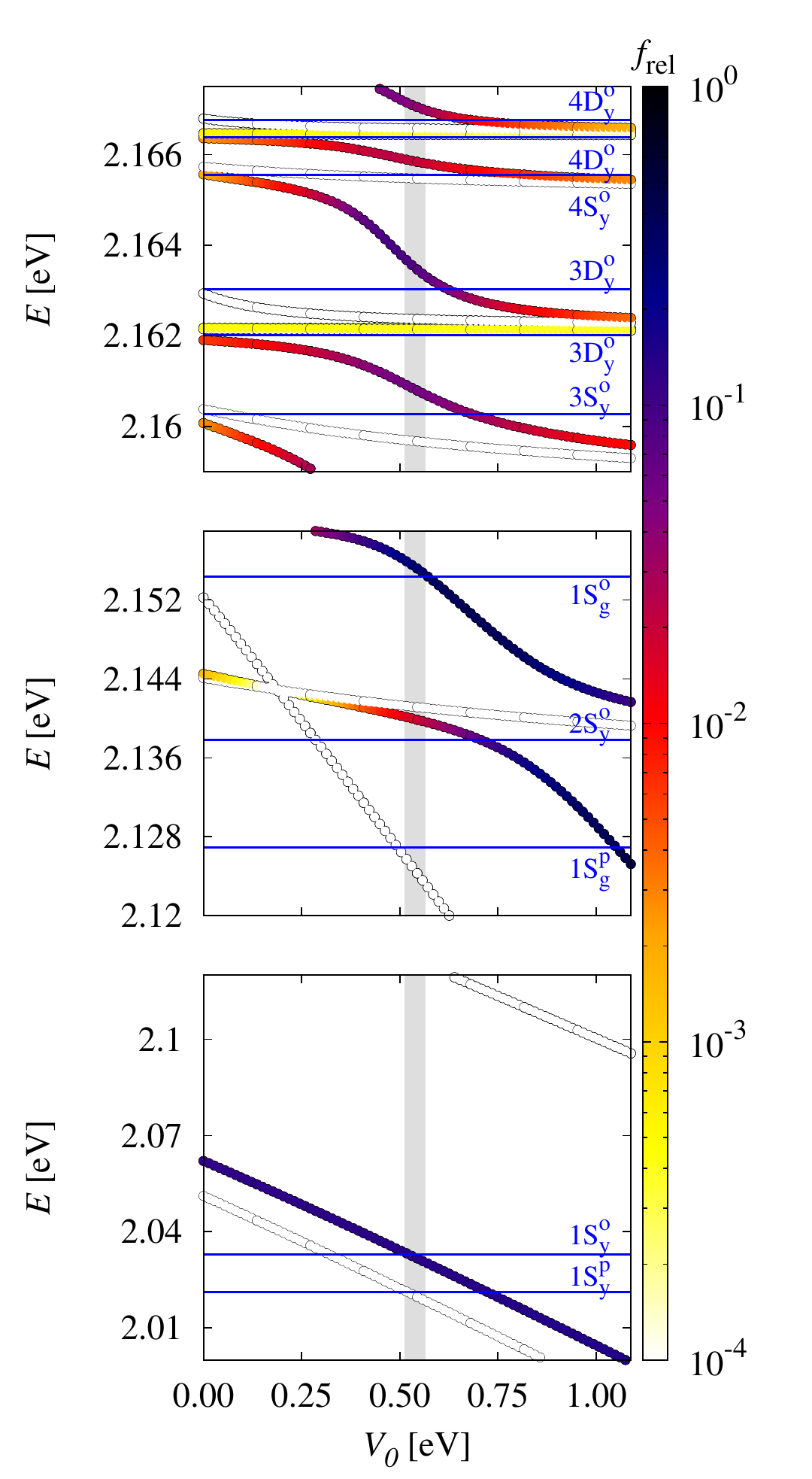}

\caption{Behavior of the
even exciton states as functions of $V_0$
when using $V_{\mathrm{CCC}}^{\mathrm{H}}$ [see Eq.~(\ref{eq:HCCC_H})].
The color bar shows the relative oscillator strengths 
for two-photon absorption.
The blue straight lines denote the position of the dipole-allowed
$\Gamma_5^+$ $S$ and $D$ exciton states observed in the experiment.
We also show the positions of the $1S$ para excitons $(1S_{\mathrm{y/g}}^{\mathrm{p}})$.
The gray area indicates the optimum range of $V_0=0.539\pm\,0.027\,\mathrm{eV}$, where
the ratio of the relative oscillator strengths of the 
yellow $2S$ and the green $1S$ state
amounts to $\sim 16$.
The effect of the central-cell corrections on the
whole even exciton spectrum is evident.
For further information see text.\label{fig:Fig4}}

\end{figure}

\begin{figure}

\includegraphics[width=1.0\columnwidth]{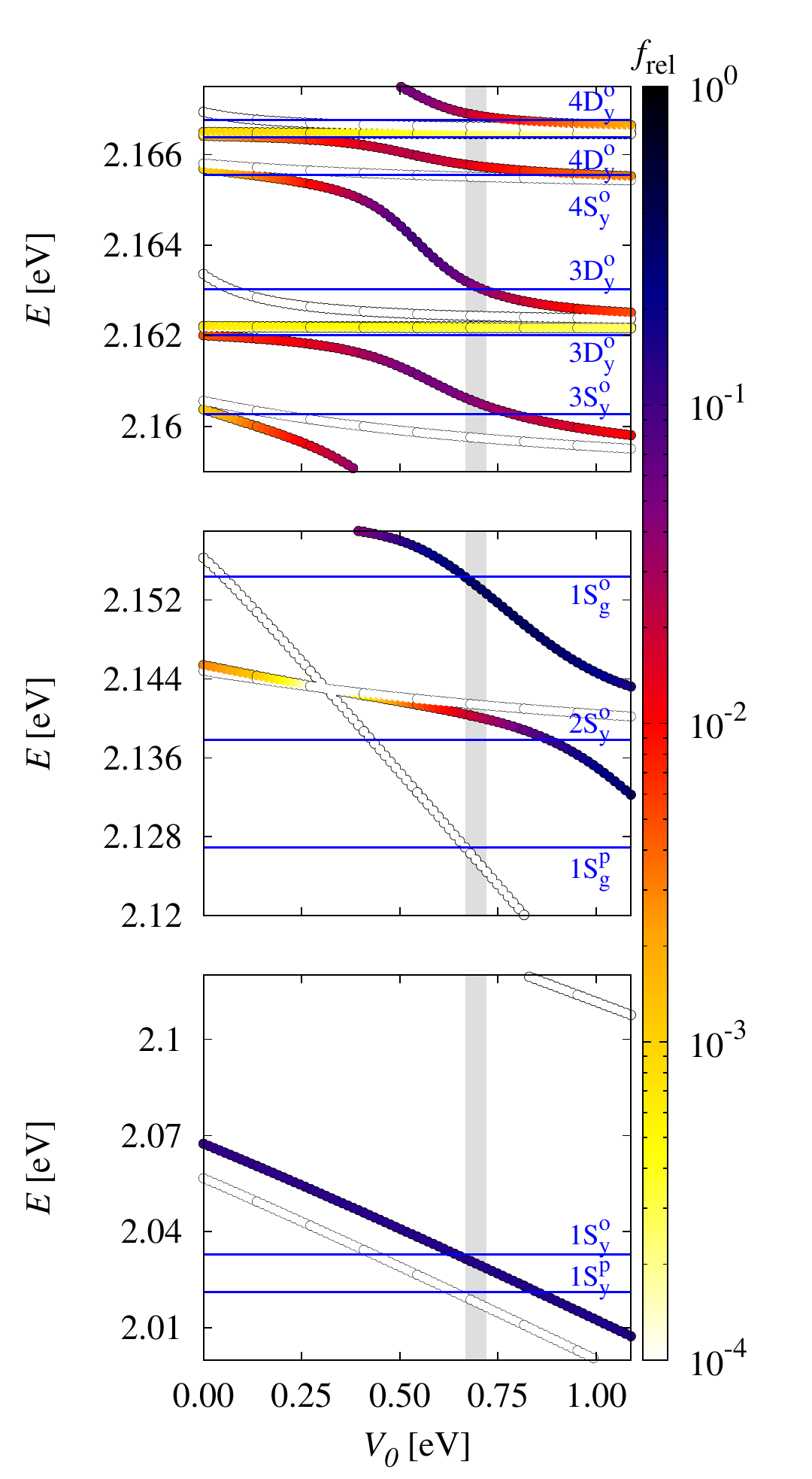}

\caption{Same calculation as in
Fig.~\ref{fig:Fig4} but with $V_{\mathrm{CCC}}^{\mathrm{PB}}$ [see Eq.~(\ref{eq:HCCC_PB})].
One can see only slight differences for the $n=1$ and $n=2$
exciton states when comparing the results to Fig.~\ref{fig:Fig4}.
The gray area indicates the optimum range of $V_0=0.694\pm\,0.027\,\mathrm{eV}$.
\label{fig:Fig5}}

\end{figure}

The figures~\ref{fig:Fig4} and~\ref{fig:Fig5} show the
effect of the correction with the coefficient $V_0$ on the spectrum
for the Haken and the Pollman-B\"uttner potential, respectivley.
As can be seen from these figures, the exchange splitting of the 
$1S$ state hardly changes when varying the value $V_0$.
Hence, we can determine
$V_0$ almost independently of $J_0$.

To find the optimum value of $V_0$, 
we compare our results to the energies of the even parity
exciton states given in Refs.~\cite{7,80,78,HO,DB_49,7_22}.
However, we can see from
Figs.~\ref{fig:Fig4} and~\ref{fig:Fig5} that there is no value
of $V_0$ for which all theoretical results
take the values of the experimentally determined
energies. This is not unexpected since the
central-cell corrections are 
only an attempt to account for the specific properties of the $1S$ exciton
within the continuum limit of Wannier excitons and are not an exact
description of this exciton state. Hence, we do not expect a perfect agreement
between theory and experiment.

Small deviations from the experimental values
could also be explained by small uncertainties in the Luttinger parameters $\gamma_i$,
$\eta_i$~\cite{80,100} or the band gap energy~\cite{GRE} as well as
by a finite temperature or small strains in the crystal. 
On the other hand, it is also possible that the experimental values 
are affected by uncertainties. This can be seen, e.g., 
when comparing the slightly different experimental results of Refs.~\cite{80} and~\cite{78,HO}. 

Note that the almost perfect agreement between theoretical and 
experimental results in Refs.~\cite{6,7} 
could only be obtained
by taking also $\gamma_1'$, $\mu'$ and $\Delta$ as fit parameters
to the experiment. However, these parameters are connected to the
band structure in $\mathrm{Cu_{2}O}$~\cite{20} and cannot 
be chosen arbitrarily~\cite{28,100}.

\begin{table}

\protect\caption{Decomposition of the irreducible representations
of the rotation group or the angular momentum states by the
cubic group $O_{\mathrm{h}}$. Note that the quasi-spin $I$ already enters the momentum $F$
via $J$. The irreducible representations
denote the symmetry of the envelope function $\left(L\right)$,
the combined symmetry of envelope and hole $\left(F\right)$
or the complete symmetry of the exciton $\left(F_t\right)$.\label{tab:2}}

\begin{centering} 
\begin{tabular}{ll|ll|ll}
\hline 
\hline
\multicolumn{2}{l}{$L$$\phantom{\int_{\frac{1}{1}}^{\frac{1}{1}}}$} & \multicolumn{2}{l}{$F=L+J\,\left(J=\frac{1}{2}\right)$ } & \multicolumn{2}{l}{$F_{t}=F+S_{\mathrm{e}}$}\tabularnewline
\hline 
 & \multicolumn{1}{l}{} &  & \multicolumn{1}{l}{} &  & \tabularnewline
\multirow{2}{*}{$0$} & \multirow{2}{*}{$\Gamma_{1}^{+}$} & \multirow{2}{*}{$\frac{1}{2}$} & \multirow{2}{*}{$\Gamma_{7}^{+}$} & $0$ & $\Gamma_{2}^{+}$\tabularnewline
 &  &  &  & $1$ & $\Gamma_{5}^{+}$\tabularnewline
 & \multicolumn{1}{l}{} &  & \multicolumn{1}{l}{} &  & \tabularnewline
\hline 
 & \multicolumn{1}{l}{} &  & \multicolumn{1}{l}{} &  & \tabularnewline
\multirow{5}{*}{$1$} & \multirow{5}{*}{$\Gamma_{4}^{-}$} & \multirow{2}{*}{$\frac{1}{2}$} & \multirow{2}{*}{$\Gamma_{7}^{-}$} & $0$ & $\Gamma_{2}^{-}$\tabularnewline
 &  &  &  & $1$ & $\Gamma_{5}^{-}$\tabularnewline
 &  &  & \multicolumn{1}{l}{} &  & \tabularnewline
 &  & \multirow{2}{*}{$\frac{3}{2}$} & \multirow{2}{*}{$\Gamma_{8}^{-}$} & $1$ & $\Gamma_{4}^{-}$\tabularnewline
 &  &  &  & $2$ & $\Gamma_{3}^{-}\oplus\Gamma_{5}^{-}$\tabularnewline
 & \multicolumn{1}{l}{} &  & \multicolumn{1}{l}{} &  & \tabularnewline
\hline 
 & \multicolumn{1}{l}{} &  & \multicolumn{1}{l}{} &  & \tabularnewline
\multirow{5}{*}{$2$} & \multirow{5}{*}{$\Gamma_{3}^{+}\oplus\Gamma_{5}^{+}$} & \multirow{2}{*}{$\frac{3}{2}$} & \multirow{2}{*}{$\Gamma_{8}^{+}$} & $1$ & $\Gamma_{5}^{+}$\tabularnewline
 &  &  &  & $2$ & $\Gamma_{3}^{+}\oplus\Gamma_{4}^{+}$\tabularnewline
 &  &  & \multicolumn{1}{l}{} &  & \tabularnewline
 &  & \multirow{2}{*}{$\frac{5}{2}$} & \multirow{2}{*}{$\Gamma_{6}^{+}\oplus\Gamma_{8}^{+}$} & $2$ & $\Gamma_{3}^{+}\oplus\Gamma_{4}^{+}$\tabularnewline
 &  &  &  & $3$ & $\Gamma_{1}^{+}\oplus\Gamma_{4}^{+}\oplus\Gamma_{5}^{+}$\tabularnewline
 & \multicolumn{1}{l}{} &  & \multicolumn{1}{l}{} &  & \tabularnewline
\hline 
 & \multicolumn{1}{l}{} &  & \multicolumn{1}{l}{} &  & \tabularnewline
\multirow{5}{*}{$3$} & \multirow{5}{*}{$\Gamma_{2}^{-}\oplus\Gamma_{4}^{-}\oplus\Gamma_{5}^{-}$} & \multirow{2}{*}{$\frac{5}{2}$} & \multirow{2}{*}{$\Gamma_{6}^{-}\oplus\Gamma_{8}^{-}$} & $2$ & $\Gamma_{3}^{-}\oplus\Gamma_{4}^{-}$\tabularnewline
 &  &  &  & $3$ & $\Gamma_{1}^{-}\oplus\Gamma_{4}^{-}\oplus\Gamma_{5}^{-}$\tabularnewline
 &  &  & \multicolumn{1}{l}{} &  & \tabularnewline
 &  & \multirow{2}{*}{$\frac{7}{2}$} & \multirow{2}{*}{$\Gamma_{6}^{-}\oplus\Gamma_{7}^{-}\oplus\Gamma_{8}^{-}$} & $3$ & $\Gamma_{1}^{-}\oplus\Gamma_{4}^{-}\oplus\Gamma_{5}^{-}$\tabularnewline
 &  &  &  & $4$ & $\Gamma_{2}^{-}\oplus\Gamma_{3}^{-}\oplus\Gamma_{4}^{-}\oplus\Gamma_{5}^{-}$\tabularnewline
 & \multicolumn{1}{l}{} &  & \multicolumn{1}{l}{} &  & \tabularnewline
\hline 
\multicolumn{2}{l}{$L$$\phantom{\int_{\frac{1}{1}}^{\frac{1}{1}}}$} & \multicolumn{2}{l}{$F=L+J\,\left(J=\frac{3}{2}\right)$ } & \multicolumn{2}{l}{$F_{t}=F+S_{\mathrm{e}}$}\tabularnewline
\hline 
 & \multicolumn{1}{l}{} &  & \multicolumn{1}{l}{} &  & \tabularnewline
\multirow{2}{*}{$0$} & \multirow{2}{*}{$\Gamma_{1}^{+}$} & \multirow{2}{*}{$\frac{3}{2}$} & \multirow{2}{*}{$\Gamma_{8}^{+}$} & $1$ & $\Gamma_{5}^{+}$\tabularnewline
 &  &  &  & $2$ & $\Gamma_{3}^{+}\oplus\Gamma_{4}^{+}$\tabularnewline
 & \multicolumn{1}{l}{} &  & \multicolumn{1}{l}{} &  & \tabularnewline
\hline 
\end{tabular}
\par\end{centering}

\end{table}

\begin{figure*}

\includegraphics[width=2.0\columnwidth]{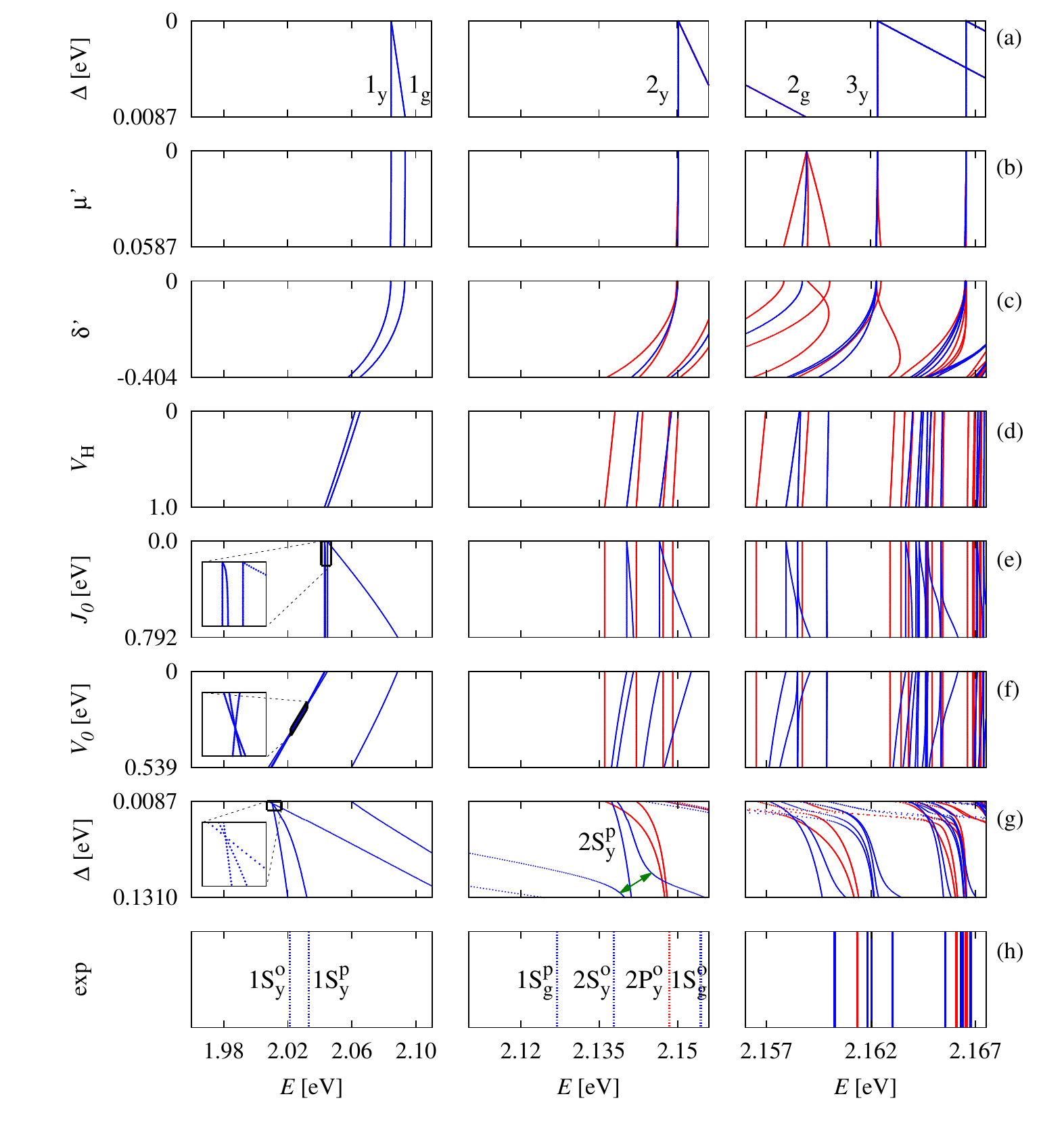}

\caption{Exciton spectrum of the even (blue)
and odd (red) exciton states when increasing all material
parameters from zero (top) to their correct values (bottom)
and using $V_{\mathrm{CCC}}^{\mathrm{H}}$ [cf. Eq.~(\ref{eq:HCCC_PB})].
If all material parameters except for $\gamma_1'$ are set to zero,
one obtains a hydrogen-like spectrum, for which the yellow (y)
and green (g) exciton states are degenerate [$\Delta=0$ in (a)]. When increasing the 
spin-orbit coupling constant $\Delta$, this degeneracy is lifted and the green
exciton states are shifted towards higher energies (a). Note that
we increase $\Delta$ in two steps to its true value of $\Delta=0.131\,\mathrm{eV}$ 
for reasons of clarity.
One can then follow these states from (b) to (g). 
Since the effect of the parameters $\eta_1'$, $\nu$ and $\tau$
on the exciton spectrum is small they are immediately set 
from zero to their correct values between (c) and (d).
In (g) and (h) the para and ortho
exciton states are denoted by an upper index p and o.
The final results at the bottom of (g), which are also listed in Table~\ref{tab:3},
can then be compared to the position of the exciton states obtained from
experiments (h).
Note that due to the marked anticrossing [green arrow in the second panel of (g)] the
assignment of the green $1S$ state and the yellow $2S$ state changes.
\label{fig:Fig6}}

\end{figure*}

\begin{table*}

\protect\caption{Comparison of calculated energies $E_{\mathrm{theor}}$ 
to experimental 
values $E_{\mathrm{exp}}$ (References given behind experimental values)
when using the central-cell corrections with the Haken potential~(\ref{eq:HCCC_H}).
The even exciton states are listed in blue text and the odd exciton states in red text.
Note that we use $E_{\mathrm{g}}=2.17202\,\mathrm{eV}$ instead of
$E_{\mathrm{g}}=2.17208\,\mathrm{eV}$~\cite{GRE} to obtain a better agreement.
The assignment of the states in the first column is motivated
by Fig.~\ref{fig:Fig6} but is generally not instructive
due to the large deviations from the hydrogen-like model.
Hence, we also give the symmetry of the states. In the case of the $P$ and $F$
excitons we do not give the symmetry of the complete exciton
state but only the combined symmetry of envelope and hole.
As regards the $5G$ excitons we only give the average energy of the
states of symmetry $\Gamma_5^+$.
The value in the fourth column gives the relative oscillator strength
in one-photon absorption ($nP$, $nF$ excitons; see Ref.~\cite{100}) or in
two-photon absorption ($nS$, $nD$, $nG$ excitons; see Eq.~(\ref{eq:frelT})).
Note that due to the interaction with the $1S_{\mathrm{g}}$ state the oscillator strength of the $2S_{\mathrm{y}}$ state
is much smaller than expected when assuming two independent, i.e., green and yellow, series.
The value in the last column indicates the percentage of the
$J=3/2$ component of the state, i.e., the green part.
Note that due to the interaction between the yellow and 
the green exciton series the green $1S$ state is spread over
several yellow exciton states. The green states with $n\geq 2$
are located far above the states listed here.
\label{tab:3}}

\begin{centering}
\begin{tabular}{cccccccc|cccccccc}
\multicolumn{2}{c}{State} &  & $E_{\mathrm{exp}}$ {[}eV{]} & $E_{\mathrm{theor}}$ {[}eV{]} & $f_{\mathrm{rel}}$ & gp {[}\%{]} &  &  & \multicolumn{2}{c}{State} &  & $E_{\mathrm{exp}}$ {[}eV{]} & $E_{\mathrm{theor}}$ {[}eV{]} & $f_{\mathrm{rel}}$ & gp {[}\%{]}\tabularnewline
\hline 
 &  &  &  &  &  &  &  &  &  &  &  &  &  &  & \tabularnewline
\textcolor{blue}{$1S_{\mathrm{y}}$} & \textcolor{blue}{$\Gamma_{2}^{+}$} & \textcolor{blue}{$\phantom{\int_{\frac{1}{1}}^{\frac{1}{1}}}$} & \textcolor{blue}{$2.0212$~\cite{DB_49}} & \textcolor{blue}{2.0200} & \textcolor{blue}{--} & \textcolor{blue}{5.49} &  &  & \textcolor{blue}{$4D_{\mathrm{y}}$} & \textcolor{blue}{$\Gamma_{3/4}^{+}$} & \textcolor{blue}{$\phantom{\int_{\frac{1}{1}}^{\frac{1}{1}}}$} & \textcolor{blue}{$2.16629$~\cite{78,HO}} & \textcolor{blue}{2.16644} & \textcolor{blue}{--} & \textcolor{blue}{0.19}\tabularnewline
\textcolor{blue}{$1S_{\mathrm{y}}$} & \textcolor{blue}{$\Gamma_{5}^{+}$} & \textcolor{blue}{$\phantom{\int_{\frac{1}{1}}^{\frac{1}{1}}}$} & \textcolor{blue}{$2.0330$~\cite{7}} & \textcolor{blue}{2.0320} & \textcolor{blue}{26.60} & \textcolor{blue}{7.22} &  &  & \textcolor{blue}{$4D_{\mathrm{y}}$} & \textcolor{blue}{$\Gamma_{5}^{+}$} & \textcolor{blue}{$\phantom{\int_{\frac{1}{1}}^{\frac{1}{1}}}$} & \textcolor{blue}{$2.16638$~\cite{78,HO}} & \textcolor{blue}{2.16645} & \textcolor{blue}{0.07} & \textcolor{blue}{0.19}\tabularnewline
 &  &  &  &  &  &  &  &  & \textcolor{blue}{$4D_{\mathrm{y}}$} & \textcolor{blue}{$\Gamma_{1/4}^{+}$} & \textcolor{blue}{$\phantom{\int_{\frac{1}{1}}^{\frac{1}{1}}}$} & \textcolor{blue}{--} & \textcolor{blue}{2.16646} & \textcolor{blue}{--} & \textcolor{blue}{0.16}\tabularnewline
\textcolor{blue}{$1S_{\mathrm{g}}$} & \textcolor{blue}{$\Gamma_{3/4}^{+}$} & \textcolor{blue}{$\phantom{\int_{\frac{1}{1}}^{\frac{1}{1}}}$} & \textcolor{blue}{$2.1269$~\cite{7_22}} & \textcolor{blue}{2.1245} & \textcolor{blue}{--} & \textcolor{blue}{71.62} &  &  & \textcolor{red}{$4F_{\mathrm{y}}$} & \textcolor{red}{$\Gamma_{7}^{-}$} & \textcolor{red}{$\phantom{\int_{\frac{1}{1}}^{\frac{1}{1}}}$} & \textcolor{red}{--} & \textcolor{red}{2.16653} & \textcolor{red}{--} & \textcolor{red}{0.12}\tabularnewline
 &  &  &  &  &  &  &  &  & \textcolor{red}{$4F_{\mathrm{y}}$} & \textcolor{red}{$\Gamma_{8}^{-}$} & \textcolor{red}{$\phantom{\int_{\frac{1}{1}}^{\frac{1}{1}}}$} & \textcolor{red}{$2.16652$~\cite{28}} & \textcolor{red}{2.16654} & \textcolor{red}{0.066} & \textcolor{red}{0.10}\tabularnewline
\textcolor{blue}{$2S_{\mathrm{y}}$} & \textcolor{blue}{$\Gamma_{5}^{+}$} & \textcolor{blue}{$\phantom{\int_{\frac{1}{1}}^{\frac{1}{1}}}$} & \textcolor{blue}{$2.1378$~\cite{7}} & \textcolor{blue}{2.1399} & \textcolor{blue}{3.55} & \textcolor{blue}{10.88} &  &  & \textcolor{red}{$4F_{\mathrm{y}}$} & \textcolor{red}{$\Gamma_{8}^{-}$} & \textcolor{red}{$\phantom{\int_{\frac{1}{1}}^{\frac{1}{1}}}$} & \textcolor{red}{$2.16654$~\cite{28}} & \textcolor{red}{2.16656} & \textcolor{red}{0.002} & \textcolor{red}{0.08}\tabularnewline
\textcolor{blue}{$2S_{\mathrm{y}}$} & \textcolor{blue}{$\Gamma_{2}^{+}$} & \textcolor{blue}{$\phantom{\int_{\frac{1}{1}}^{\frac{1}{1}}}$} & \textcolor{blue}{--} & \textcolor{blue}{2.1412} & \textcolor{blue}{--} & \textcolor{blue}{1.43} &  &  & \textcolor{red}{$4F_{\mathrm{y}}$} & \textcolor{red}{$\Gamma_{6}^{-}$} & \textcolor{red}{$\phantom{\int_{\frac{1}{1}}^{\frac{1}{1}}}$} & \textcolor{red}{$2.16654$~\cite{28}} & \textcolor{red}{2.16657} & \textcolor{red}{0.010} & \textcolor{red}{0.08}\tabularnewline
\textcolor{red}{$2P_{\mathrm{y}}$} & \textcolor{red}{$\Gamma_{8}^{-}$} & \textcolor{red}{$\phantom{\int_{\frac{1}{1}}^{\frac{1}{1}}}$} & \textcolor{red}{$2.1484$~\cite{GRE}} & \textcolor{red}{2.1475} & \textcolor{red}{351.4} & \textcolor{red}{1.91} &  &  & \textcolor{red}{$4F_{\mathrm{y}}$} & \textcolor{red}{$\Gamma_{6}^{-}$} & \textcolor{red}{$\phantom{\int_{\frac{1}{1}}^{\frac{1}{1}}}$} & \textcolor{red}{$2.16658$~\cite{28}} & \textcolor{red}{2.16660} & \textcolor{red}{0.011} & \textcolor{red}{0.06}\tabularnewline
\textcolor{red}{$2P_{\mathrm{y}}$} & \textcolor{red}{$\Gamma_{7}^{-}$} & \textcolor{red}{$\phantom{\int_{\frac{1}{1}}^{\frac{1}{1}}}$} & \textcolor{red}{--} & \textcolor{red}{2.1480} & \textcolor{red}{--} & \textcolor{red}{1.30} &  &  & \textcolor{blue}{$4D_{\mathrm{y}}$} & \textcolor{blue}{$\Gamma_{3/4}^{+}$} & \textcolor{blue}{$\phantom{\int_{\frac{1}{1}}^{\frac{1}{1}}}$} & \textcolor{blue}{--} & \textcolor{blue}{2.16658} & \textcolor{blue}{--} & \textcolor{blue}{0.22}\tabularnewline
 &  &  &  &  &  &  &  &  & \textcolor{blue}{$4D_{\mathrm{y}}$} & \textcolor{blue}{$\Gamma_{5}^{+}$} & \textcolor{blue}{$\phantom{\int_{\frac{1}{1}}^{\frac{1}{1}}}$} & \textcolor{blue}{$2.16677$~\cite{78,HO}} & \textcolor{blue}{2.16704} & \textcolor{blue}{6.86} & \textcolor{blue}{3.67}\tabularnewline
\textcolor{blue}{$1S_{\mathrm{g}}$} & \textcolor{blue}{$\Gamma_{5}^{+}$} & \textcolor{blue}{$\phantom{\int_{\frac{1}{1}}^{\frac{1}{1}}}$} & \textcolor{blue}{$2.1544$~\cite{7}} & \textcolor{blue}{2.1553} & \textcolor{blue}{56.01} & \textcolor{blue}{36.88} &  &  &  &  &  &  &  &  & \tabularnewline
 &  &  &  &  &  &  &  &  & \textcolor{blue}{$5S_{\mathrm{y}}$} & \textcolor{blue}{$\Gamma_{2}^{+}$} & \textcolor{blue}{$\phantom{\int_{\frac{1}{1}}^{\frac{1}{1}}}$} & \textcolor{blue}{--} & \textcolor{blue}{2.16798} & \textcolor{blue}{--} & \textcolor{blue}{0.10}\tabularnewline
\textcolor{blue}{$3S_{\mathrm{y}}$} & \textcolor{blue}{$\Gamma_{2}^{+}$} & \textcolor{blue}{$\phantom{\int_{\frac{1}{1}}^{\frac{1}{1}}}$} & \textcolor{blue}{--} & \textcolor{blue}{2.15967} & \textcolor{blue}{--} & \textcolor{blue}{0.48} &  &  & \textcolor{blue}{$5S_{\mathrm{y}}$} & \textcolor{blue}{$\Gamma_{5}^{+}$} & \textcolor{blue}{$\phantom{\int_{\frac{1}{1}}^{\frac{1}{1}}}$} & \textcolor{blue}{$2.16801$~\cite{78,HO}} & \textcolor{blue}{2.16816} & \textcolor{blue}{2.02} & \textcolor{blue}{0.81}\tabularnewline
\textcolor{blue}{$3S_{\mathrm{y}}$} & \textcolor{blue}{$\Gamma_{5}^{+}$} & \textcolor{blue}{$\phantom{\int_{\frac{1}{1}}^{\frac{1}{1}}}$} & \textcolor{blue}{$2.16027$~\cite{78,HO}} & \textcolor{blue}{2.16080} & \textcolor{blue}{10.34} & \textcolor{blue}{4.49} &  &  & \textcolor{red}{$5P_{\mathrm{y}}$} & \textcolor{red}{$\Gamma_{8}^{-}$} & \textcolor{red}{$\phantom{\int_{\frac{1}{1}}^{\frac{1}{1}}}$} & \textcolor{red}{$2.16829$~\cite{GRE}} & \textcolor{red}{2.16825} & \textcolor{red}{32.82} & \textcolor{red}{0.25}\tabularnewline
\textcolor{red}{$3P_{\mathrm{y}}$} & \textcolor{red}{$\Gamma_{8}^{-}$} & \textcolor{red}{$\phantom{\int_{\frac{1}{1}}^{\frac{1}{1}}}$} & \textcolor{red}{$2.16135$~\cite{GRE}} & \textcolor{red}{2.16119} & \textcolor{red}{147.3} & \textcolor{red}{0.93} &  &  & \textcolor{red}{$5P_{\mathrm{y}}$} & \textcolor{red}{$\Gamma_{7}^{-}$} & \textcolor{red}{$\phantom{\int_{\frac{1}{1}}^{\frac{1}{1}}}$} & \textcolor{red}{--} & \textcolor{red}{2.16830} & \textcolor{red}{--} & \textcolor{red}{0.18}\tabularnewline
\textcolor{red}{$3P_{\mathrm{y}}$} & \textcolor{red}{$\Gamma_{7}^{-}$} & \textcolor{red}{$\phantom{\int_{\frac{1}{1}}^{\frac{1}{1}}}$} & \textcolor{red}{--} & \textcolor{red}{2.16141} & \textcolor{red}{--} & \textcolor{red}{0.63} &  &  & \textcolor{blue}{$5D_{\mathrm{y}}$} & \textcolor{blue}{$\Gamma_{3/4}^{+}$} & \textcolor{blue}{$\phantom{\int_{\frac{1}{1}}^{\frac{1}{1}}}$} & \textcolor{blue}{$2.16841$~\cite{78,HO}} & \textcolor{blue}{2.16846} & \textcolor{blue}{--} & \textcolor{blue}{0.11}\tabularnewline
\textcolor{blue}{$3D_{\mathrm{y}}$} & \textcolor{blue}{$\Gamma_{3/4}^{+}$} & \textcolor{blue}{$\phantom{\int_{\frac{1}{1}}^{\frac{1}{1}}}$} & \textcolor{blue}{$2.16183$~\cite{78,HO}} & \textcolor{blue}{2.16213} & \textcolor{blue}{--} & \textcolor{blue}{0.31} &  &  & \textcolor{blue}{$5D_{\mathrm{y}}$} & \textcolor{blue}{$\Gamma_{5}^{+}$} & \textcolor{blue}{$\phantom{\int_{\frac{1}{1}}^{\frac{1}{1}}}$} & \textcolor{blue}{$2.16846$~\cite{78,HO}} & \textcolor{blue}{2.16846} & \textcolor{blue}{0.05} & \textcolor{blue}{0.12}\tabularnewline
\textcolor{blue}{$3D_{\mathrm{y}}$} & \textcolor{blue}{$\Gamma_{5}^{+}$} & \textcolor{blue}{$\phantom{\int_{\frac{1}{1}}^{\frac{1}{1}}}$} & \textcolor{blue}{$2.16202$~\cite{78,HO}} & \textcolor{blue}{2.16215} & \textcolor{blue}{0.09} & \textcolor{blue}{0.30} &  &  & \textcolor{blue}{$5D_{\mathrm{y}}$} & \textcolor{blue}{$\Gamma_{1/4}^{+}$} & \textcolor{blue}{$\phantom{\int_{\frac{1}{1}}^{\frac{1}{1}}}$} & \textcolor{blue}{--} & \textcolor{blue}{2.16847} & \textcolor{blue}{--} & \textcolor{blue}{0.10}\tabularnewline
\textcolor{blue}{$3D_{\mathrm{y}}$} & \textcolor{blue}{$\Gamma_{1/4}^{+}$} & \textcolor{blue}{$\phantom{\int_{\frac{1}{1}}^{\frac{1}{1}}}$} & \textcolor{blue}{--} & \textcolor{blue}{2.16217} & \textcolor{blue}{--} & \textcolor{blue}{0.25} &  &  & \textcolor{red}{$5F_{\mathrm{y}}$} & \textcolor{red}{$\Gamma_{7}^{-}$} & \textcolor{red}{$\phantom{\int_{\frac{1}{1}}^{\frac{1}{1}}}$} & \textcolor{red}{--} & \textcolor{red}{2.16850} & \textcolor{red}{--} & \textcolor{red}{0.09}\tabularnewline
\textcolor{blue}{$3D_{\mathrm{y}}$} & \textcolor{blue}{$\Gamma_{3/4}^{+}$} & \textcolor{blue}{$\phantom{\int_{\frac{1}{1}}^{\frac{1}{1}}}$} & \textcolor{blue}{--} & \textcolor{blue}{2.16237} & \textcolor{blue}{--} & \textcolor{blue}{0.46} &  &  & \textcolor{red}{$5F_{\mathrm{y}}$} & \textcolor{red}{$\Gamma_{8}^{-}$} & \textcolor{red}{$\phantom{\int_{\frac{1}{1}}^{\frac{1}{1}}}$} & \textcolor{red}{$2.16851$~\cite{28}} & \textcolor{red}{2.16850} & \textcolor{red}{0.069} & \textcolor{red}{0.07}\tabularnewline
\textcolor{blue}{$3D_{\mathrm{y}}$} & \textcolor{blue}{$\Gamma_{5}^{+}$} & \textcolor{blue}{$\phantom{\int_{\frac{1}{1}}^{\frac{1}{1}}}$} & \textcolor{blue}{$2.16303$~\cite{78,HO}} & \textcolor{blue}{2.16348} & \textcolor{blue}{15.04} & \textcolor{blue}{8.49} &  &  & \textcolor{red}{$5F_{\mathrm{y}}$} & \textcolor{red}{$\Gamma_{8}^{-}$} & \textcolor{red}{$\phantom{\int_{\frac{1}{1}}^{\frac{1}{1}}}$} & \textcolor{red}{$2.16852$~\cite{28}} & \textcolor{red}{2.16852} & \textcolor{red}{0.000} & \textcolor{red}{0.06}\tabularnewline
 &  &  &  &  &  &  &  &  & \textcolor{red}{$5F_{\mathrm{y}}$} & \textcolor{red}{$\Gamma_{6}^{-}$} & \textcolor{red}{$\phantom{\int_{\frac{1}{1}}^{\frac{1}{1}}}$} & \textcolor{red}{$2.16852$~\cite{28}} & \textcolor{red}{2.16852} & \textcolor{red}{0.002} & \textcolor{red}{0.06}\tabularnewline
\textcolor{blue}{$4S_{\mathrm{y}}$} & \textcolor{blue}{$\Gamma_{2}^{+}$} & \textcolor{blue}{$\phantom{\int_{\frac{1}{1}}^{\frac{1}{1}}}$} & \textcolor{blue}{--} & \textcolor{blue}{2.16547} & \textcolor{blue}{--} & \textcolor{blue}{0.21} &  &  & \textcolor{red}{$5F_{\mathrm{y}}$} & \textcolor{red}{$\Gamma_{6}^{-}$} & \textcolor{red}{$\phantom{\int_{\frac{1}{1}}^{\frac{1}{1}}}$} & \textcolor{red}{$2.16855$~\cite{28}} & \textcolor{red}{2.16855} & \textcolor{red}{0.001} & \textcolor{red}{0.04}\tabularnewline
\textcolor{blue}{$4S_{\mathrm{y}}$} & \textcolor{blue}{$\Gamma_{5}^{+}$} & \textcolor{blue}{$\phantom{\int_{\frac{1}{1}}^{\frac{1}{1}}}$} & \textcolor{blue}{$2.16555$~\cite{78,HO}} & \textcolor{blue}{2.16584} & \textcolor{blue}{3.79} & \textcolor{blue}{1.53} &  &  & \textcolor{blue}{$5D_{\mathrm{y}}$} & \textcolor{blue}{$\Gamma_{3/4}^{+}$} & \textcolor{blue}{$\phantom{\int_{\frac{1}{1}}^{\frac{1}{1}}}$} & \textcolor{blue}{--} & \textcolor{blue}{2.16854} & \textcolor{blue}{--} & \textcolor{blue}{0.11}\tabularnewline
\textcolor{red}{$4P_{\mathrm{y}}$} & \textcolor{red}{$\Gamma_{8}^{-}$} & \textcolor{red}{$\phantom{\int_{\frac{1}{1}}^{\frac{1}{1}}}$} & \textcolor{red}{$2.16609$~\cite{GRE}} & \textcolor{red}{2.16604} & \textcolor{red}{67.43} & \textcolor{red}{0.45} &  &  & \textcolor{blue}{$5\bar{G}_{\mathrm{y}}$} & \textcolor{blue}{$\Gamma_{5}^{+}$} & \textcolor{blue}{$\phantom{\int_{\frac{1}{1}}^{\frac{1}{1}}}$} & \textcolor{blue}{--} & \textcolor{blue}{2.16855} & \textcolor{blue}{0.00} & \textcolor{blue}{0.03}\tabularnewline
\textcolor{red}{$4P_{\mathrm{y}}$} & \textcolor{red}{$\Gamma_{7}^{-}$} & \textcolor{red}{$\phantom{\int_{\frac{1}{1}}^{\frac{1}{1}}}$} & \textcolor{red}{--} & \textcolor{red}{2.16614} & \textcolor{red}{--} & \textcolor{red}{0.32} &  &  & \textcolor{blue}{$5D_{\mathrm{y}}$} & \textcolor{blue}{$\Gamma_{5}^{+}$} & \textcolor{blue}{$\phantom{\int_{\frac{1}{1}}^{\frac{1}{1}}}$} & \textcolor{blue}{$2.16860$~\cite{78,HO}} & \textcolor{blue}{2.16879} & \textcolor{blue}{4.30} & \textcolor{blue}{2.22}\tabularnewline
\hline 
\end{tabular}
\par\end{centering}

\end{table*}

\begin{table*}

\protect\caption{Same comparison as in Table~\ref{tab:3} but 
when using the central-cell corrections with the Pollmann-B\"uttner potential~(\ref{eq:HCCC_PB}).
Especially for the states with $n<3$ differences in the calculated energies
can be observed when using the different corrections~(\ref{eq:HCCC_H}) or~(\ref{eq:HCCC_PB}).
Note that for each $n$ the relative oscillator strength of one $nD$ state is larger than the
relative oscillator strengths of the $nS$ state in accordance with the experimental 
results of Ref.~\cite{7}.
\label{tab:4}}

\begin{centering}
\begin{tabular}{cccccccc|cccccccc}
\multicolumn{2}{c}{State} &  & $E_{\mathrm{exp}}$ {[}eV{]} & $E_{\mathrm{theor}}$ {[}eV{]} & $f_{\mathrm{rel}}$ & gp {[}\%{]} &  &  & \multicolumn{2}{c}{State} &  & $E_{\mathrm{exp}}$ {[}eV{]} & $E_{\mathrm{theor}}$ {[}eV{]} & $f_{\mathrm{rel}}$ & gp {[}\%{]}\tabularnewline
\hline 
 &  &  &  &  &  &  &  &  &  &  &  &  &  &  & \tabularnewline
\textcolor{blue}{$1S_{\mathrm{y}}$} & \textcolor{blue}{$\Gamma_{2}^{+}$} & \textcolor{blue}{$\phantom{\int_{\frac{1}{1}}^{\frac{1}{1}}}$} & \textcolor{blue}{$2.0212$~\cite{DB_49}} & \textcolor{blue}{2.0180} & \textcolor{blue}{--} & \textcolor{blue}{5.49} &  &  & \textcolor{blue}{$4D_{\mathrm{y}}$} & \textcolor{blue}{$\Gamma_{3/4}^{+}$} & \textcolor{blue}{$\phantom{\int_{\frac{1}{1}}^{\frac{1}{1}}}$} & \textcolor{blue}{$2.16629$~\cite{78,HO}} & \textcolor{blue}{2.16646} & \textcolor{blue}{--} & \textcolor{blue}{0.17}\tabularnewline
\textcolor{blue}{$1S_{\mathrm{y}}$} & \textcolor{blue}{$\Gamma_{5}^{+}$} & \textcolor{blue}{$\phantom{\int_{\frac{1}{1}}^{\frac{1}{1}}}$} & \textcolor{blue}{$2.0330$~\cite{7}} & \textcolor{blue}{2.0300} & \textcolor{blue}{27.90} & \textcolor{blue}{6.83} &  &  & \textcolor{blue}{$4D_{\mathrm{y}}$} & \textcolor{blue}{$\Gamma_{5}^{+}$} & \textcolor{blue}{$\phantom{\int_{\frac{1}{1}}^{\frac{1}{1}}}$} & \textcolor{blue}{$2.16638$~\cite{78,HO}} & \textcolor{blue}{2.16647} & \textcolor{blue}{0.53} & \textcolor{blue}{0.18}\tabularnewline
 &  &  &  &  &  &  &  &  & \textcolor{blue}{$4D_{\mathrm{y}}$} & \textcolor{blue}{$\Gamma_{1/4}^{+}$} & \textcolor{blue}{$\phantom{\int_{\frac{1}{1}}^{\frac{1}{1}}}$} & \textcolor{blue}{--} & \textcolor{blue}{2.16648} & \textcolor{blue}{--} & \textcolor{blue}{0.15}\tabularnewline
\textcolor{blue}{$1S_{\mathrm{g}}$} & \textcolor{blue}{$\Gamma_{3/4}^{+}$} & \textcolor{blue}{$\phantom{\int_{\frac{1}{1}}^{\frac{1}{1}}}$} & \textcolor{blue}{$2.1269$~\cite{7_22}} & \textcolor{blue}{2.1254} & \textcolor{blue}{--} & \textcolor{blue}{65.53} &  &  & \textcolor{red}{$4F_{\mathrm{y}}$} & \textcolor{red}{$\Gamma_{7}^{-}$} & \textcolor{red}{$\phantom{\int_{\frac{1}{1}}^{\frac{1}{1}}}$} & \textcolor{red}{--} & \textcolor{red}{2.16653} & \textcolor{red}{--} & \textcolor{red}{0.12}\tabularnewline
 &  &  &  &  &  &  &  &  & \textcolor{red}{$4F_{\mathrm{y}}$} & \textcolor{red}{$\Gamma_{8}^{-}$} & \textcolor{red}{$\phantom{\int_{\frac{1}{1}}^{\frac{1}{1}}}$} & \textcolor{red}{$2.16652$~\cite{28}} & \textcolor{red}{2.16654} & \textcolor{red}{0.078} & \textcolor{red}{0.10}\tabularnewline
\textcolor{blue}{$2S_{\mathrm{y}}$} & \textcolor{blue}{$\Gamma_{5}^{+}$} & \textcolor{blue}{$\phantom{\int_{\frac{1}{1}}^{\frac{1}{1}}}$} & \textcolor{blue}{$2.1378$~\cite{7}} & \textcolor{blue}{2.1401} & \textcolor{blue}{4.22} & \textcolor{blue}{11.16} &  &  & \textcolor{red}{$4F_{\mathrm{y}}$} & \textcolor{red}{$\Gamma_{8}^{-}$} & \textcolor{red}{$\phantom{\int_{\frac{1}{1}}^{\frac{1}{1}}}$} & \textcolor{red}{$2.16654$~\cite{28}} & \textcolor{red}{2.16657} & \textcolor{red}{0.002} & \textcolor{red}{0.08}\tabularnewline
\textcolor{blue}{$2S_{\mathrm{y}}$} & \textcolor{blue}{$\Gamma_{2}^{+}$} & \textcolor{blue}{$\phantom{\int_{\frac{1}{1}}^{\frac{1}{1}}}$} & \textcolor{blue}{--} & \textcolor{blue}{2.1414} & \textcolor{blue}{--} & \textcolor{blue}{1.31} &  &  & \textcolor{red}{$4F_{\mathrm{y}}$} & \textcolor{red}{$\Gamma_{6}^{-}$} & \textcolor{red}{$\phantom{\int_{\frac{1}{1}}^{\frac{1}{1}}}$} & \textcolor{red}{$2.16654$~\cite{28}} & \textcolor{red}{2.16657} & \textcolor{red}{0.009} & \textcolor{red}{0.08}\tabularnewline
\textcolor{red}{$2P_{\mathrm{y}}$} & \textcolor{red}{$\Gamma_{8}^{-}$} & \textcolor{red}{$\phantom{\int_{\frac{1}{1}}^{\frac{1}{1}}}$} & \textcolor{red}{$2.1484$~\cite{GRE}} & \textcolor{red}{2.1482} & \textcolor{red}{292.3} & \textcolor{red}{1.72} &  &  & \textcolor{red}{$4F_{\mathrm{y}}$} & \textcolor{red}{$\Gamma_{6}^{-}$} & \textcolor{red}{$\phantom{\int_{\frac{1}{1}}^{\frac{1}{1}}}$} & \textcolor{red}{$2.16658$~\cite{28}} & \textcolor{red}{2.16660} & \textcolor{red}{0.010} & \textcolor{red}{0.05}\tabularnewline
\textcolor{red}{$2P_{\mathrm{y}}$} & \textcolor{red}{$\Gamma_{7}^{-}$} & \textcolor{red}{$\phantom{\int_{\frac{1}{1}}^{\frac{1}{1}}}$} & \textcolor{red}{--} & \textcolor{red}{2.1486} & \textcolor{red}{--} & \textcolor{red}{1.20} &  &  & \textcolor{blue}{$4D_{\mathrm{y}}$} & \textcolor{blue}{$\Gamma_{3/4}^{+}$} & \textcolor{blue}{$\phantom{\int_{\frac{1}{1}}^{\frac{1}{1}}}$} & \textcolor{blue}{--} & \textcolor{blue}{2.16661} & \textcolor{blue}{--} & \textcolor{blue}{0.19}\tabularnewline
 &  &  &  &  &  &  &  &  & \textcolor{blue}{$4D_{\mathrm{y}}$} & \textcolor{blue}{$\Gamma_{5}^{+}$} & \textcolor{blue}{$\phantom{\int_{\frac{1}{1}}^{\frac{1}{1}}}$} & \textcolor{blue}{$2.16677$~\cite{78,HO}} & \textcolor{blue}{2.16686} & \textcolor{blue}{3.24} & \textcolor{blue}{1.82}\tabularnewline
\textcolor{blue}{$1S_{\mathrm{g}}$} & \textcolor{blue}{$\Gamma_{5}^{+}$} & \textcolor{blue}{$\phantom{\int_{\frac{1}{1}}^{\frac{1}{1}}}$} & \textcolor{blue}{$2.1544$~\cite{7}} & \textcolor{blue}{2.1535} & \textcolor{blue}{65.25} & \textcolor{blue}{42.41} &  &  &  &  &  &  &  &  & \tabularnewline
 &  &  &  &  &  &  &  &  & \textcolor{blue}{$5S_{\mathrm{y}}$} & \textcolor{blue}{$\Gamma_{2}^{+}$} & \textcolor{blue}{$\phantom{\int_{\frac{1}{1}}^{\frac{1}{1}}}$} & \textcolor{blue}{--} & \textcolor{blue}{2.16800} & \textcolor{blue}{--} & \textcolor{blue}{0.09}\tabularnewline
\textcolor{blue}{$3S_{\mathrm{y}}$} & \textcolor{blue}{$\Gamma_{2}^{+}$} & \textcolor{blue}{$\phantom{\int_{\frac{1}{1}}^{\frac{1}{1}}}$} & \textcolor{blue}{--} & \textcolor{blue}{2.15974} & \textcolor{blue}{--} & \textcolor{blue}{0.44} &  &  & \textcolor{blue}{$5S_{\mathrm{y}}$} & \textcolor{blue}{$\Gamma_{5}^{+}$} & \textcolor{blue}{$\phantom{\int_{\frac{1}{1}}^{\frac{1}{1}}}$} & \textcolor{blue}{$2.16801$~\cite{78,HO}} & \textcolor{blue}{2.16811} & \textcolor{blue}{1.17} & \textcolor{blue}{0.48}\tabularnewline
\textcolor{blue}{$3S_{\mathrm{y}}$} & \textcolor{blue}{$\Gamma_{5}^{+}$} & \textcolor{blue}{$\phantom{\int_{\frac{1}{1}}^{\frac{1}{1}}}$} & \textcolor{blue}{$2.16027$~\cite{78,HO}} & \textcolor{blue}{2.16053} & \textcolor{blue}{7.83} & \textcolor{blue}{3.29} &  &  & \textcolor{red}{$5P_{\mathrm{y}}$} & \textcolor{red}{$\Gamma_{8}^{-}$} & \textcolor{red}{$\phantom{\int_{\frac{1}{1}}^{\frac{1}{1}}}$} & \textcolor{red}{$2.16829$~\cite{GRE}} & \textcolor{red}{2.16829} & \textcolor{red}{28.17} & \textcolor{red}{0.24}\tabularnewline
\textcolor{red}{$3P_{\mathrm{y}}$} & \textcolor{red}{$\Gamma_{8}^{-}$} & \textcolor{red}{$\phantom{\int_{\frac{1}{1}}^{\frac{1}{1}}}$} & \textcolor{red}{$2.16135$~\cite{GRE}} & \textcolor{red}{2.16138} & \textcolor{red}{125.9} & \textcolor{red}{0.86} &  &  & \textcolor{red}{$5P_{\mathrm{y}}$} & \textcolor{red}{$\Gamma_{7}^{-}$} & \textcolor{red}{$\phantom{\int_{\frac{1}{1}}^{\frac{1}{1}}}$} & \textcolor{red}{--} & \textcolor{red}{2.16834} & \textcolor{red}{--} & \textcolor{red}{0.17}\tabularnewline
\textcolor{red}{$3P_{\mathrm{y}}$} & \textcolor{red}{$\Gamma_{7}^{-}$} & \textcolor{red}{$\phantom{\int_{\frac{1}{1}}^{\frac{1}{1}}}$} & \textcolor{red}{--} & \textcolor{red}{2.16158} & \textcolor{red}{--} & \textcolor{red}{0.60} &  &  & \textcolor{blue}{$5D_{\mathrm{y}}$} & \textcolor{blue}{$\Gamma_{3/4}^{+}$} & \textcolor{blue}{$\phantom{\int_{\frac{1}{1}}^{\frac{1}{1}}}$} & \textcolor{blue}{$2.16841$~\cite{78,HO}} & \textcolor{blue}{2.16847} & \textcolor{blue}{--} & \textcolor{blue}{0.10}\tabularnewline
\textcolor{blue}{$3D_{\mathrm{y}}$} & \textcolor{blue}{$\Gamma_{3/4}^{+}$} & \textcolor{blue}{$\phantom{\int_{\frac{1}{1}}^{\frac{1}{1}}}$} & \textcolor{blue}{$2.16183$~\cite{78,HO}} & \textcolor{blue}{2.16217} & \textcolor{blue}{--} & \textcolor{blue}{0.28} &  &  & \textcolor{blue}{$5D_{\mathrm{y}}$} & \textcolor{blue}{$\Gamma_{5}^{+}$} & \textcolor{blue}{$\phantom{\int_{\frac{1}{1}}^{\frac{1}{1}}}$} & \textcolor{blue}{$2.16846$~\cite{78,HO}} & \textcolor{blue}{2.16847} & \textcolor{blue}{0.04} & \textcolor{blue}{0.12}\tabularnewline
\textcolor{blue}{$3D_{\mathrm{y}}$} & \textcolor{blue}{$\Gamma_{5}^{+}$} & \textcolor{blue}{$\phantom{\int_{\frac{1}{1}}^{\frac{1}{1}}}$} & \textcolor{blue}{$2.16202$~\cite{78,HO}} & \textcolor{blue}{2.16219} & \textcolor{blue}{0.07} & \textcolor{blue}{0.29} &  &  & \textcolor{blue}{$5D_{\mathrm{y}}$} & \textcolor{blue}{$\Gamma_{1/4}^{+}$} & \textcolor{blue}{$\phantom{\int_{\frac{1}{1}}^{\frac{1}{1}}}$} & \textcolor{blue}{--} & \textcolor{blue}{2.16848} & \textcolor{blue}{--} & \textcolor{blue}{0.09}\tabularnewline
\textcolor{blue}{$3D_{\mathrm{y}}$} & \textcolor{blue}{$\Gamma_{1/4}^{+}$} & \textcolor{blue}{$\phantom{\int_{\frac{1}{1}}^{\frac{1}{1}}}$} & \textcolor{blue}{--} & \textcolor{blue}{2.16221} & \textcolor{blue}{--} & \textcolor{blue}{0.24} &  &  & \textcolor{red}{$5F_{\mathrm{y}}$} & \textcolor{red}{$\Gamma_{7}^{-}$} & \textcolor{red}{$\phantom{\int_{\frac{1}{1}}^{\frac{1}{1}}}$} & \textcolor{red}{--} & \textcolor{red}{2.16850} & \textcolor{red}{--} & \textcolor{red}{0.09}\tabularnewline
\textcolor{blue}{$3D_{\mathrm{y}}$} & \textcolor{blue}{$\Gamma_{3/4}^{+}$} & \textcolor{blue}{$\phantom{\int_{\frac{1}{1}}^{\frac{1}{1}}}$} & \textcolor{blue}{--} & \textcolor{blue}{2.16243} & \textcolor{blue}{--} & \textcolor{blue}{0.41} &  &  & \textcolor{red}{$5F_{\mathrm{y}}$} & \textcolor{red}{$\Gamma_{8}^{-}$} & \textcolor{red}{$\phantom{\int_{\frac{1}{1}}^{\frac{1}{1}}}$} & \textcolor{red}{$2.16851$~\cite{28}} & \textcolor{red}{2.16851} & \textcolor{red}{0.078} & \textcolor{red}{0.07}\tabularnewline
\textcolor{blue}{$3D_{\mathrm{y}}$} & \textcolor{blue}{$\Gamma_{5}^{+}$} & \textcolor{blue}{$\phantom{\int_{\frac{1}{1}}^{\frac{1}{1}}}$} & \textcolor{blue}{$2.16303$~\cite{78,HO}} & \textcolor{blue}{2.16308} & \textcolor{blue}{8.42} & \textcolor{blue}{4.87} &  &  & \textcolor{red}{$5F_{\mathrm{y}}$} & \textcolor{red}{$\Gamma_{8}^{-}$} & \textcolor{red}{$\phantom{\int_{\frac{1}{1}}^{\frac{1}{1}}}$} & \textcolor{red}{$2.16852$~\cite{28}} & \textcolor{red}{2.16852} & \textcolor{red}{0.000} & \textcolor{red}{0.06}\tabularnewline
 &  &  &  &  &  &  &  &  & \textcolor{red}{$5F_{\mathrm{y}}$} & \textcolor{red}{$\Gamma_{6}^{-}$} & \textcolor{red}{$\phantom{\int_{\frac{1}{1}}^{\frac{1}{1}}}$} & \textcolor{red}{$2.16852$~\cite{28}} & \textcolor{red}{2.16853} & \textcolor{red}{0.001} & \textcolor{red}{0.06}\tabularnewline
\textcolor{blue}{$4S_{\mathrm{y}}$} & \textcolor{blue}{$\Gamma_{2}^{+}$} & \textcolor{blue}{$\phantom{\int_{\frac{1}{1}}^{\frac{1}{1}}}$} & \textcolor{blue}{--} & \textcolor{blue}{2.16550} & \textcolor{blue}{--} & \textcolor{blue}{0.19} &  &  & \textcolor{red}{$5F_{\mathrm{y}}$} & \textcolor{red}{$\Gamma_{6}^{-}$} & \textcolor{red}{$\phantom{\int_{\frac{1}{1}}^{\frac{1}{1}}}$} & \textcolor{red}{$2.16855$~\cite{28}} & \textcolor{red}{2.16855} & \textcolor{red}{0.001} & \textcolor{red}{0.04}\tabularnewline
\textcolor{blue}{$4S_{\mathrm{y}}$} & \textcolor{blue}{$\Gamma_{5}^{+}$} & \textcolor{blue}{$\phantom{\int_{\frac{1}{1}}^{\frac{1}{1}}}$} & \textcolor{blue}{$2.16555$~\cite{78,HO}} & \textcolor{blue}{2.16575} & \textcolor{blue}{2.45} & \textcolor{blue}{0.98} &  &  & \textcolor{blue}{$5D_{\mathrm{y}}$} & \textcolor{blue}{$\Gamma_{3/4}^{+}$} & \textcolor{blue}{$\phantom{\int_{\frac{1}{1}}^{\frac{1}{1}}}$} & \textcolor{blue}{--} & \textcolor{blue}{2.16855} & \textcolor{blue}{0.00} & \textcolor{blue}{0.03}\tabularnewline
\textcolor{red}{$4P_{\mathrm{y}}$} & \textcolor{red}{$\Gamma_{8}^{-}$} & \textcolor{red}{$\phantom{\int_{\frac{1}{1}}^{\frac{1}{1}}}$} & \textcolor{red}{$2.16609$~\cite{GRE}} & \textcolor{red}{2.16612} & \textcolor{red}{58.29} & \textcolor{red}{0.43} &  &  & \textcolor{blue}{$5\bar{G}_{\mathrm{y}}$} & \textcolor{blue}{$\Gamma_{5}^{+}$} & \textcolor{blue}{$\phantom{\int_{\frac{1}{1}}^{\frac{1}{1}}}$} & \textcolor{blue}{--} & \textcolor{blue}{2.16856} & \textcolor{blue}{--} & \textcolor{blue}{0.07}\tabularnewline
\textcolor{red}{$4P_{\mathrm{y}}$} & \textcolor{red}{$\Gamma_{7}^{-}$} & \textcolor{red}{$\phantom{\int_{\frac{1}{1}}^{\frac{1}{1}}}$} & \textcolor{red}{--} & \textcolor{red}{2.16621} & \textcolor{red}{--} & \textcolor{red}{0.31} &  &  & \textcolor{blue}{$5D_{\mathrm{y}}$} & \textcolor{blue}{$\Gamma_{5}^{+}$} & \textcolor{blue}{$\phantom{\int_{\frac{1}{1}}^{\frac{1}{1}}}$} & \textcolor{blue}{$2.16860$~\cite{78,HO}} & \textcolor{blue}{2.16868} & \textcolor{blue}{1.68} & \textcolor{blue}{0.92}\tabularnewline
\hline 
\end{tabular}
\par\end{centering}

\end{table*}

It can be seen from Figs.~\ref{fig:Fig4} and~\ref{fig:Fig5} that the oscillator
strength of the exciton state at $E\approx 2.143\,\mathrm{eV}$
changes rapidly with increasing $V_0$. 
From the experimental results of Refs.~\cite{6,7} we know
that the two exciton states at $E=2.1378\,\mathrm{eV}$ and $E=2.1544\,\mathrm{eV}$
are well separated from the other exciton states and
that the phonon background is small.
Hence, the ratio of the relative two-photon oscillator strengths
can be calculated quite accurately to $\sim\!16$. 

We now choose the value of $V_0$ such 
that the ratio of the calculated two-photon oscillator strengths 
reaches the same value and obtain
\begin{equation}
V_0=0.539\pm\,0.027\,\mathrm{eV}\label{eq:V0H}
\end{equation}
when using the Haken potential [cf. Eq.~(\ref{eq:HCCC_H})] or
\begin{equation}
V_0=0.694\pm\,0.027\,\mathrm{eV}\label{eq:V0PB}
\end{equation}
when using the Pollmann-B\"uttner potential [cf. Eq.~(\ref{eq:HCCC_PB})].
Note that the error bars for $V_0$ are chosen such
that the ratio of the oscillator strengths lies between $14$ and $18$.

Having determined the most suitable values of
$V_0$ and $J_0$, we can now turn our attention to the exciton
Bohr radius $a_{\mathrm{exc}}^{\left(1S\right)}$ of the $1S$ ortho exciton and to the
correct assignment of the $n=2$ exciton states.

To determine the radius $a_{\mathrm{exc}}^{\left(1S\right)}$,
we evaluate 
\begin{eqnarray}
\left\langle\Psi\middle| r\middle|\Psi\right\rangle & = & \sum_{N'}\sum_{NLJFF_t M_{F_t}}c_{N'LJFF_t M_{F_t}}c_{NLJFF_t M_{F_t}}\nonumber\\
\nonumber\\
& \times & \sum_{j=-2}^{2}\frac{\alpha\left(R_2 \right)^{j}_{NL}}{N+L+j+1}\delta_{N',\,N+j}\label{eq:PsirPsi}
\end{eqnarray}
with the wave function $\Psi$ of Eq.~(\ref{eq:ansatz}) 
and compare the result with the formula~\cite{GRE_1}
\begin{equation}
\left\langle r\right\rangle=\frac{1}{2}a_{\mathrm{exc}}\left[3n^2-L\left(L+1\right)\right]\label{eq:aexcH}
\end{equation}
known from the hydrogen atom, where we set $n=1$ and $L=0$.
Note that the function $\left(R_2 \right)^{j}_{NL}$ in Eq.~(\ref{eq:PsirPsi}) 
is taken from the recursion relations of the Coulomb-Sturmian
functions in the Appendix of Ref.~\cite{100}.
We obtain
\begin{equation}
a_{\mathrm{exc}}^{\left(1S\right)}\approx 0.793\,\mathrm{nm}\approx 1.86\,a 
\end{equation}
when using the Haken potential or
\begin{equation}
a_{\mathrm{exc}}^{\left(1S\right)}\approx 0.810\,\mathrm{nm}\approx 1.90\,a
\end{equation}
when using the Pollmann-B\"uttner potential.
In both cases the radius of the $1S$ ortho exciton is
large enough that the corrections to the kinetic energy discussed
in Sec.~\ref{sec:Bandstruc} can certainly be neglected.

Let us now proceed to the correct assignment of the $n=2$ exciton states.
Since in the investigation of Uihlein~\emph{et~al}~\cite{6,7} the 
wrong values for the Luttinger parameters were used (cf.~Ref.~\cite{100}),
it is not clear whether the state at $E=2.1544\,\mathrm{eV}$ can still be assigned
as the yellow $2S$ ortho exciton state and the state at $E=2.1378\,\mathrm{eV}$
as the green $1S$ ortho exciton state when using the correct Luttinger parameters.

To demonstrate from which hydrogen-like states
the experimentally observed exciton states originate,
we find it instructive to start from the hydrogen-like
spectrum with almost all material parameters set to zero
and then increase these material parameters successively to their
true values. 
This is shown in Fig.~\ref{fig:Fig6}.

At first all material parameters except for $\gamma_1'$ are set to zero,
so that a true hydrogen-like spectrum is obtained, where the yellow (y)
and green (g) exciton states are degenerate. This spectrum is shown
in the panel (a) of Fig.~\ref{fig:Fig6}.
When increasing the spin-orbit coupling constant $\Delta$ in Fig.~\ref{fig:Fig6}(a),
the degeneracy between the green and the yellow exciton series is lifted.
The increase of the Luttinger parameters $\mu'$ and $\delta'$ in the panels (b) and (c)
furthermore lifts the degeneracy between the exciton states of different angular momentum $L$.
The Haken potential does not change degeneracies but slightly
lowers the energy of the exciton states in Fig.~\ref{fig:Fig6}(d).
The exchange energy described by the constant $J_0$ lifts the degenercy between ortho and
para exciton states in Fig.~\ref{fig:Fig6}(e).
As the operator $\delta\!\left(\boldsymbol{r}\right)$ affects only the states
of even parity (blue lines), the energy of the
odd exciton states (red lines) remains unchanged in Fig.~\ref{fig:Fig6}(f).
Note that we increase $\Delta$ in two steps to its true value of $\Delta=0.131\,\mathrm{eV}$ 
for reasons of clarity. Hence, at the bottom of Fig.~\ref{fig:Fig6}(g)
all material values have been increased to their true values.
For a comparison, we show in panel (h) the position of the experimentally
observed states.
Following the exciton states from panel (a) to (g), it is possible to
assign them with the notation $nL^{\mathrm{p/o}}_{\mathrm{y/g}}$,
where the upper index denotes a para or an ortho
exciton state and the lower index a yellow or a green state.

The results presented in Fig.~\ref{fig:Fig6} suggest to
assign the exciton state at $E=2.1378\,\mathrm{eV}$ to the green
$1S$ ortho exciton state.
However, one can observe an anticrossing between the green $1S$ state
and the yellow $2S$ state, which is indicated by a
green arrow in Fig.~\ref{fig:Fig6}(g).
Hence, the assignment has to be changed.
As a proof, we can calculate the percentage of the
$J=3/2$ component of these states, i.e., their green part,
by evaluating 
\begin{equation}
\mathrm{gp}=\left\langle\Psi\middle| P\middle|\Psi\right\rangle
\end{equation}
with the projection operator
\begin{equation}
P=\sum_{M_J=-3/2}^{3/2}\left|\frac{3}{2},\,M_J\right\rangle\left\langle\frac{3}{2},\,M_J\right|
\end{equation}
and the exciton wave function $\left|\Psi\right\rangle$ (see also Appendix~\ref{sub:green-part}).

The green part gp of the state at $E\approx 2.1544\,\mathrm{eV}$ is distinctly higher 
($\mathrm{gp}\approx 40\%$) than the green part of the exciton state at 
$E\approx 2.1378\,\mathrm{eV}$ ($\mathrm{gp}\approx 11\%$). However, since also $\mathrm{gp}\approx 40\%$
is significantly smaller than one, we see that the assignment
of this exciton state as the ground state of the green series is
questionable and shows the significant deviations from the
hydrogen-like model.
The green $1S$ exciton state is distributed over the yellow states.
Note that in Ref.~\cite{7} also the state of higher energy
had a larger green part than the state of lower energy. 
However, in Fig.~2 of Ref.~\cite{7} the assignment 
is reversed since the limit of
$\mu'\rightarrow 0$ was used to designate the states.
It seems obvious that a similar anticrossing
between the green $1S$ state
and the yellow $2S$ state was disregarded.
A considerable effect of the interaction
between the green and yellow series is the change in the oscillator
strength of the states. The oscillator strength of the $2S_y$ state
is much smaller than expected when assuming two independent, i.e., 
green and yellow, series~\cite{6,7} (cf. also Tables~\ref{tab:3} and~\ref{tab:4}).

For reasons of completeness, we
give the size of the green $1S$ and the yellow $2S$ state by evaluating Eq.~(\ref{eq:PsirPsi}).
Since these states are strongly mixed and a correct assignment with a principal quantum
number $n$ is not possible, we do not use the formula~(\ref{eq:aexcH}). We obtain
\begin{subequations}
\begin{eqnarray}
\left\langle r\right\rangle\left(2S_y\right) & \approx & 4.32\,\mathrm{nm}\approx 10.1\,a,\\
\left\langle r\right\rangle\left(1S_g\right) & \approx & 5.32\,\mathrm{nm}\approx 12.5\,a,
\end{eqnarray}
\end{subequations}
when using the Haken potential or
\begin{subequations}
\begin{eqnarray}
\left\langle r\right\rangle\left(2S_y\right) & \approx & 4.39\,\mathrm{nm}\approx 10.3\,a,\\
\left\langle r\right\rangle\left(1S_g\right) & \approx & 4.09\,\mathrm{nm}\approx 9.58\,a,
\end{eqnarray}
\end{subequations}
when using the Pollmann-B\"uttner potential.
We see that in both cases the values of $\left\langle r\right\rangle$ 
for the green $1S$ and the yellow $2S$ state are of the same size.
This is expected due to the strong mixing of both states.

The resonance of the green $1S$ state with the 
yellow exciton series and the mixing of all even exciton
states via the cubic band structure leads to an admixture
of $D$ and $G$ states to the green $1S$ state. Hence, 
the three $\Gamma_5^+$ states which we assigned with $1S_{\mathrm{g}}$
are elliptically deformed and invariant only under the subgroup
$D_{4\mathrm{h}}$ of $O_{\mathrm{h}}$~\cite{100,G3}.
The lower symmetry of the envelope function allows
for a smaller mean distance between electron and hole in
a specific direction, which leads to a gain of energy due to
the Coulomb interaction~\cite{100}.
As regards the $xy$-component, the symmetry axis 
of the according subgroup $D_{4\mathrm{h}}$ is the $z$-axis of the crystal.
Since for this state the expectation values
$\left\langle\Psi\middle| x^2\middle|\Psi\right\rangle$ and
$\left\langle\Psi\middle| y^2\middle|\Psi\right\rangle$ are identical,
we can calculate the semi-principal axes of the elliptically
deformed state by evaluating
\begin{eqnarray}
\left\langle\Psi\middle| x^2\middle|\Psi\right\rangle & = & \left\langle\Psi\middle|\frac{1}{2}\left(r^2-z^2\right)\middle|\Psi\right\rangle\nonumber\\
& = & \sum_{N'L'J'F'F_t' M_{F_t}'}\sum_{NLJFF_t M_{F_t}}\nonumber\\
\nonumber\\
 & & c_{N'L'J'F'F_t' M_{F_t}'}c_{NLJFF_t M_{F_t}}\nonumber\\
\nonumber\\
& \times & \alpha^2\left\langle\Pi'\middle|\frac{1}{3}r^2-\frac{1}{3\sqrt{6}}X^{(2)}_0\middle|\Pi\right\rangle
\label{eq:Psix2Psi}
\end{eqnarray}
and
\begin{eqnarray}
\left\langle\Psi\middle| z^2\middle|\Psi\right\rangle & = & \sum_{N'L'J'F'F_t' M_{F_t}'}\sum_{NLJFF_t M_{F_t}}\nonumber\\
\nonumber\\
 & & c_{N'L'J'F'F_t' M_{F_t}'}c_{NLJFF_t M_{F_t}}\nonumber\\
\nonumber\\
& \times & \alpha^2\left\langle\Pi'\middle|\frac{1}{3}\sqrt{\frac{2}{3}}X^{(2)}_0+\frac{1}{3}r^2\middle|\Pi\right\rangle
\label{eq:Psix2Psi}
\end{eqnarray}
with the wave function $\Psi$ of Eq.~(\ref{eq:ansatz}) 
and the matrix elements $\langle\Pi'|X^{(2)}_0|\Pi\rangle$
and $\langle\Pi'|r^2|\Pi\rangle$
listed in the Appendix of Ref.~\cite{125}.
We obtain
\begin{eqnarray}
\left\langle x^2\right\rangle & \approx & 116.4\,a^2,\nonumber\\
\left\langle z^2\right\rangle & \approx & 29.9\,a^2,
\end{eqnarray}
when using the Haken potential or
\begin{eqnarray}
\left\langle x^2\right\rangle & \approx & 68.6\,a^2,\nonumber\\
\left\langle z^2\right\rangle & \approx & 25.1\,a^2,
\end{eqnarray}
when using the Pollmann-B\"uttner potential.
The significant differences in $\left\langle x^2\right\rangle$
and $\left\langle z^2\right\rangle$ show again the
strong resonance of the green $1S$ state with the 
yellow series as well as the strong admixture of states
with $L\geq 2$. 
We finally want to note that, due to the
coupling of the yellow and green series, the green $1S$ has to be regarded
as an excited state in the complete exciton spectrum and not as
the ground state of the green series. In particular, the
green $1S$ state is orthogonal to the true ground state of the complete spectrum,
i.e., to the yellow $1S$ state.

Let us now discuss the other exciton states.
To determine the number of para and ortho exciton states as well as their 
degeneracies for the different values of $L$,
one can use group theoretical considerations.
In the spherical approximation, in which the cubic
part of the Hamiltonian is neglected $\left(\delta'=0\right)$, the momentum $F=J+L$ is a good quantum
number for the states of negative parity since the exchange interaction does
not act on these states. The states of positive parity can be classified by
the total momentum $F_t=F+S_{\mathrm{e}}$ in the spherical
approximation.

If the complete cubic Hamiltonian is treated, the reduction
of the irreducible representations $D^{F}$ or $D^{F_t}$ of the rotation group
by the cubic group $O_{\mathrm{h}}$ has to be considered~\cite{G1}. 
This is shown in Table~\ref{tab:2}.
As has already been stated in Ref.~\cite{100},
a normal spin one transforms according to the irreducible representation
$\Gamma_{4}^{+}$ of the cubic group whereas the quasi-spin $I$ transforms
according to $\Gamma_{5}^{+}=\Gamma_{4}^{+}\otimes\Gamma_{2}^{+}$. Therefore,
one has to include the additional factor $\Gamma_{2}^{+}$ when determining
the symmetry of an exciton state~\cite{7,28,100}. This symmetry is given by the symmetry
of the envelope function, the valence band, and the conduction
band:
\begin{equation}
\Gamma_{\mathrm{exc}}=\Gamma_{\mathrm{env}}\otimes\Gamma_{\mathrm{v}}\otimes\Gamma_{\mathrm{c}}.
\end{equation}
Only states of symmetry $\Gamma_4^-$ are dipole allowed in one-photon
absorption and only states of symmetry $\Gamma_5^+$ are dipole allowed in two-photon
absorption.
Hence, we see from Table~\ref{tab:2} that there
are at the most one $P$ state and four $F$ states
or one $S$ and two $D$ states
for each principal quantum number $n$,
which can be observed in experiments.

Since the exchange interaction does not
act on the exciton states with negative parity,
one can use the irreducible representations
of the second column of Table~\ref{tab:2}
to classify these exciton states~\cite{28}.
For the exciton states of positive parity the
irreducible representations
of the third column are needed.
Note that the cubic part of the Hamiltonian
mixes the $S$ and $D$ exciton states of symmetry $\Gamma_5^+$.
Hence, the exchange interaction acts only on the $D$
excitons of symmetry $\Gamma_5^+$ via their $S$ component.
The degeneracies between the $D$ states
of symmetry $\Gamma_3^+$ and $\Gamma_4^+$ or
$\Gamma_1^+$ and $\Gamma_4^+$ is not lifted, respectively 
(cf.~the third column of Table~\ref{tab:2}).

Since neither $J$ nor $F$ are good quantum numbers
due to the cubic symmetry of our Hamiltonian, we do not
use the nomenclature $n^{2J+1}L_{F}$ of Refs.~\cite{6,7}.
Although $L$ is likewise
no good quantum number, the assignment of the exciton
states by using $S$, $P$, $D$, $F$ and $G$ to denote
the angular momentum is still common
(see, e.g., Refs.~\cite{80,28}). Hence, we feel obliged to
classify the states by introducing the notation 
$nL_{\mathrm{y}/\mathrm{g}}$ for comparison with other works
but also stress that this is generally not instructive
due to the large deviations from the hydrogen-like model
(cf.~also Ref.~\cite{100}).
By the index y or g we denote the yellow or the green exciton series,
respectively.
To be more correct, we will also give the symmetry of the exciton states 
in terms of the irreducible representations
of Table~\ref{tab:2}. These symmetries can be 
determined by regarding the eigenvectors of the
generalized eigenvalue problem~(\ref{eq:gev})~\cite{100}.

In the Tables~\ref{tab:3} and~\ref{tab:4} we now give
a direct comparison between experimental and theoretical exciton
energies for all states with $n\leq 5$. 
One can see that the results with the Haken potential
listed in Tab.~\ref{tab:3} show a better agreement with the experimental values
than the results with the Pollmann-B\"uttner potential
listed in Tab.~\ref{tab:4}.
Hence, we have chosen the central-cell corrections with the 
Haken potential for the calculation of Fig.~\ref{fig:Fig6}.

The Haken potential or the Pollmann-B\"uttner
potential also slightly affects the odd exciton series and especially the $2P$ exciton state.
These potentials shift the energy of the $\Gamma_4^-$ (resp. $\Gamma_8^-$) $2P$ exciton
state by an amount of $210\,\mathrm{\upmu eV}$~(Haken) or
$880\,\mathrm{\upmu eV}$~(Pollmann-B\"uttner) towards lower energies.

\section{Summary and outlook~\label{sec:Summary-and-outlook}}

We have treated the exciton spectrum of
$\mathrm{Cu_{2}O}$ considering the complete valence band structure,
the exchange interaction, and the central-cell corrections.
A thorough discussion of the central-cell corrections
revealed that only the frequency and momentum dependence of the
dielectric function $\varepsilon\left(k,\,\omega\right)$
have to be accounted for. Due to the estimated size of the
$1S$ exciton Bohr radius, corrections to the kinetic energy
can be neglected.
Hence, only the two parameters $V_{0}$ and $J_{0}$
are decisive for the relative position of the exciton states.
While $J_0$ describes the splitting of the exciton states into ortho
and para components, $V_0$ changes the relative energy of the states but
leaves this splitting between ortho and para component of the same exciton
state almost unchanged. Hence, these parameters 
could be determined almost independently.
This means that our results are not very sensitive 
to the choice of the parameters used. Instead, there is only one
combination of both parameters $J_0$ and $V_0$ given in Eqs.~(\ref{eq:J0})-(\ref{eq:V0PB}), 
for which our results are in good agreement with the experiment.

We have shown that the central-cell corrections considerably affect
the complete even exciton series since the
valence band structure couples the $1S$ state to higher exciton states.
The frequency dependence
of the dielectric function also slightly affects the odd exciton series and
lowers, in particular, the energy of the $2P$ exciton state.
Furthermore, we have demonstrated that due to the coupling of the yellow and the green exciton series
the green $1S$ exciton state is distributed over all yellow states.

In contrast to earlier works~\cite{78}, we have presented a closed
theory of the complete exciton series in $\mathrm{Cu_{2}O}$,
where we explicitly give the 
correction potentials~(\ref{eq:HCCC_H})
or~(\ref{eq:HCCC_PB}).
Hence, the introduction of quantum defects 
or the introduction of different exchange parameters for different 
exciton states, which take the effect of the central cell corrections
into account only phenomenologically, is redundant~\cite{78,80}.

The results of our theory show a very good agreement with 
experimental values (see Table.~\ref{tab:3}).
Therefore, we are confident that 
an according extension of our theory 
will allow for the calculation of exciton spectra in $\mathrm{Cu_{2}O}$
in electric or in combined electric and
magnetic fields.

\acknowledgments
We thank J. Heck\"otter, M. Bayer, D. Fr\"ohlich, M. A{\ss}mann, and
D.~Dizdarevic for helpful discussions.

\appendix

\section{$p^4$-Terms \label{sec:p4}}

As has already been stated in Sec.~\ref{sec:Bandstruc},
the terms of the fourth power of $\boldsymbol{p}$
span a fifteen dimensional space with the basis functions
\begin{equation}
p_i^4,\quad p_i^3 p_j,\quad p_i^2 p_j^2,\quad p_i p_j p_k^2
\end{equation}
with $i,j,k\in\{1,2,3\}$ and $i\neq j\neq k\neq i$.
The six linear combinations of $p^4$ terms (including the quasi spin $I$),
which transform according to $\Gamma_1^+$~\cite{G3} read in terms of irreducible tensors
\begin{subequations}
\begin{align}
\mathrm{(I):}\quad   & p^4, \\
\displaybreak[2]
\mathrm{(II):}\quad  & P^{(4)}\left(\Gamma_1^+\right), \\
\displaybreak[2]
\mathrm{(III):}\quad & p^2\left(P^{(2)}\cdot I^{(2)}\right), \\
\displaybreak[2]
\mathrm{(IV):}\quad  & p^2\left[P^{(2)}\times I^{(2)}\right]^{(4)}\left(\Gamma_1^+\right), \\
\displaybreak[2]
\mathrm{(V):}\quad   & \left[P^{(4)}\times I^{(2)}\right]^{(4)}\left(\Gamma_1^+\right), \\
\displaybreak[2]
\mathrm{(VI):}\quad  & \left[P^{(4)}\times I^{(2)}\right]^{(6)}\left(\Gamma_1^+\right),
\end{align}
\end{subequations}
with
\begin{subequations}
\begin{equation}
T^{(4)}\left(\Gamma_1^+\right)=\sqrt{\frac{5}{24}}\sum_{k=\pm 4}T^{(4)}_{k}+\sqrt{\frac{7}{12}}T^{(4)}_{0},
\end{equation}
and
\begin{equation}
T^{(6)}\left(\Gamma_1^+\right)=-\frac{\sqrt{7}}{4}\sum_{k=\pm 4}T^{(6)}_{k}+\frac{1}{\sqrt{8}}T^{(6)}_{0}.
\end{equation}
\end{subequations}

One can choose appropriate linear combinations of the states (I)-(VI):
\begin{widetext}
\begin{subequations}
\begin{align}
\frac{1}{5}\mathrm{(I)}-\frac{1}{3\sqrt{30}}\mathrm{(II)} = &\: \left[p_1^2 p_2^2 +\mathrm{c.p.}\right]
\\
\displaybreak[1]
\nonumber \\
\frac{2}{3}\hbar^2 \mathrm{(I)}+\frac{2}{45}\mathrm{(III)}+\frac{1}{18}\sqrt{\frac{24}{5}}\mathrm{(IV)} = &\: \boldsymbol{p}^2\left[p_1^2 \boldsymbol{I}_1^2+\mathrm{c.p.}\right]
\\
\displaybreak[1]
\nonumber \\
\frac{1}{30}\mathrm{(III)}-\frac{1}{36}\sqrt{\frac{24}{5}}\mathrm{(IV)} = &\: \boldsymbol{p}^2\left[p_1 p_2\left\{\boldsymbol{I}_1,\,\boldsymbol{I}_2\right\}+\mathrm{c.p.}\right]
\\
\displaybreak[1]
\nonumber \\
\frac{6}{5}\hbar^2 \mathrm{(I)}-\frac{8}{9\sqrt{30}}\hbar^2 \mathrm{(II)}-\frac{4}{27}\sqrt{\frac{7}{11}}\mathrm{(V)}+\frac{1}{9}\sqrt{\frac{14}{33}}\mathrm{(VI)} = &\: \left[\left(p_1^4+6p_2^2 p_3^2\right) \boldsymbol{I}_1^2+\mathrm{c.p.}\right]
\\
\displaybreak[1]
\nonumber \\
-\frac{1}{18}\sqrt{\frac{7}{11}}\mathrm{(V)}-\frac{1}{9}\sqrt{\frac{14}{33}}\mathrm{(IV)} = &\: \left[\left(p_1^2+ p_2^2 -6p_3^2\right)p_1 p_2 \left\{\boldsymbol{I}_1,\,\boldsymbol{I}_2\right\}+\mathrm{c.p.}\right]
\end{align}
\end{subequations}
\end{widetext}
with $\left\{ a,b\right\} =\frac{1}{2}\left(ab+ba\right)$ and
c.p.~denoting cyclic permutation. 
These linear combinations enter the generalized expressions
of the kinetic energy of the hole and the electron in Sec.~\ref{sec:Bandstruc}.

\section{Oscillator strengths \label{sub:Oscillator-strengths}}

We now give the formula for the expression
\begin{equation}
\lim_{r\rightarrow0}\,_{T}\left\langle 1,\,M'_{F_t}\middle|\Psi\left(\boldsymbol{r}\right)\right\rangle,
\end{equation}
which is needed for the evaluation of the relative 
oscillator strength $f_{\mathrm{rel}}$~(\ref{eq:frelT})
in two photon absorption experiments.
Using the wave function of Eq.~(\ref{eq:ansatz}), we find
\begin{align}
&\: \lim_{r\rightarrow0}\,_{T}\left\langle 1,\,M'_{F_t}\middle|\Psi\left(\boldsymbol{r}\right)\right\rangle\nonumber \\
= &\: \sum_{NFF_{t}}\sum_{M_{S_{\mathrm{e}}}}c_{N0FFF_{t}M'_{F_t}}\;\sqrt{\frac{2}{\alpha^3}}\nonumber \\
\displaybreak[2]
\nonumber \\
\times  &\:  \left(-1\right)^{F-2M_{S_{\mathrm{e}}}+\frac{1}{2}}\left[(2F+1)(2F_{t}+1)\right]^{\frac{1}{2}}\nonumber \\
\displaybreak[2]
\nonumber \\
\times  &\: \left(\begin{array}{ccc}
F & \frac{1}{2} & F_{t}\\
M'_{F_t}-M_{S_{\mathrm{e}}} & M_{S_{\mathrm{e}}} & -M'_{F_t}
\end{array}\right)\nonumber \\\
\displaybreak[2]
\nonumber \\
\times  &\:  \left(\begin{array}{ccc}
1 & \frac{1}{2} & F\\
M'_{F_t} & -M_{S_{\mathrm{e}}} & M_{S_{\mathrm{e}}}-M'_{F_t}
\end{array}\right).
\end{align}

\section{Green part of $\Psi$ \label{sub:green-part}}

Here we give the formula for the scalar product
which is needed to calculate the green part of the
wave function $\Psi$ as
\begin{equation}
\mathrm{gp}=\sum_{M_J=-3/2}^{3/2}\left\langle\Psi\middle|\frac{3}{2},\,M_J\right\rangle\left\langle\frac{3}{2},\,M_J\middle|\Psi\right\rangle.
\end{equation} 
We find
\begin{align}
 &\:  \left\langle\Psi\middle|\frac{3}{2},\,M_J\right\rangle\left\langle\frac{3}{2},\,M_J\middle|\Psi\right\rangle\nonumber \\
\displaybreak[2]
\nonumber \\
= &\:  \sum_{j=-1}^{1}\sum_{NLFF_{t}M_{F_t}}\sum_{F'F_{t}'M_{F_t}'}\sum_{M_{S_{\mathrm{e}}}M_L}\frac{\left(R_{1}\right)_{NL}^{j}}{N+L+j+1}\nonumber \\
\displaybreak[2]
\nonumber \\
\times  &\:  c_{(N+j)LJF'F_{t}'M_{F_t}'}c_{NLJFF_{t}M_{F_t}}\nonumber \\
\displaybreak[2]
\nonumber \\
\times  &\:  \left(-1\right)^{F+F'+M_{F_t}+M_{F_t}'-2J+2M_J-1}\nonumber \\
\displaybreak[2]
\nonumber \\
\times  &\:  \left[(2F+1)(2F_{t}+1)(2F'+1)(2F_{t}'+1)\right]^{\frac{1}{2}}\nonumber \\
\displaybreak[2]
\nonumber \\
\times  &\: \left(\begin{array}{ccc}
F & \frac{1}{2} & F_{t}\\
M_L+M_J & M_{S_{\mathrm{e}}} & -M_{F_t}
\end{array}\right)\nonumber \\
\displaybreak[2]
\nonumber \\
\times  &\: \left(\begin{array}{ccc}
L & J & F\\
M_L & M_J & -M_L-M_J
\end{array}\right)\nonumber \\
\displaybreak[2]
\nonumber \\
\times  &\: \left(\begin{array}{ccc}
F' & \frac{1}{2} & F_{t}'\\
M_L+M_J & M_{S_{\mathrm{e}}} & -M_{F_t}'
\end{array}\right)\nonumber \\
\displaybreak[2]
\nonumber \\
\times  &\: \left(\begin{array}{ccc}
L & J & F'\\
M_L & M_J & -M_L-M_J
\end{array}\right).
\end{align}

The function $\left(R_{1}\right)_{NL}^{j}$
is taken from the recursion relations of the Coulomb-Sturmian functions
in the Appendix of Ref.~\cite{100}.

\section{Matrix elements \label{sub:Matrix-elements}}

In this section we give the matrix elements of the terms in Eq.~(\ref{eq:HCCC})
in the basis of Eq.~(\ref{eq:basis}) in Hartree units. The normalization factor $N_{NL}^{\left(\alpha\right)}$
is given in Eq.~(\ref{eq:normaliz}). 
All other matrix elements, which enter the symmetric matrices $\boldsymbol{D}$ and
$\boldsymbol{M}$ in Eq.~(\ref{eq:gev}) and which are not given here, are listed in the Appendix of Ref.~\cite{100}.

\begin{widetext}

\begin{align}
\left\langle \Pi'\left|\delta\left(\boldsymbol{r}\right)\right|\Pi\right\rangle =  &\: \delta_{L'0}\delta_{L0}\delta_{JJ'}\delta_{F_{t}F'_{t}}\delta_{M_{F_{t}}M'_{F_{t}}}\:\frac{1}{\pi}\left(-1\right)^{F_{t}+F'+F+J+\frac{1}{2}}\nonumber \\
\nonumber \\
\times &\: \left[\left(2F_{t}+1\right)\left(2F+1\right)\left(2F'+1\right)\right]^{\frac{1}{2}}\left\{ \begin{array}{ccc}
F' & F_{t} & \frac{1}{2}\\
F_{t} & F & 0
\end{array}\right\} \left\{ \begin{array}{ccc}
0 & F' & J\\
F & 0 & 0
\end{array}\right\}, \label{eq:matdelta}
\\
\displaybreak[1]
\nonumber \\
\left\langle \Pi'\left|\boldsymbol{S}_{\mathrm{e}}\cdot\boldsymbol{S}_{\mathrm{h}}\,\delta\left(\boldsymbol{r}\right)\right|\Pi\right\rangle =  &\: \delta_{L'0}\delta_{L0}\delta_{F_{t}F'_{t}}\delta_{M_{F_{t}}M'_{F_{t}}}\:\frac{3}{2\pi}\left(-1\right)^{F_t+F'+F+J+J'}\nonumber \\
\nonumber \\
\times &\: \left[\left(2F+1\right)\left(2F'+1\right)\left(2J+1\right)\left(2J'+1\right)\right]^{\frac{1}{2}}\nonumber \\
\nonumber \\
\times &\: \left\{ \begin{array}{ccc}
F' & F & 1\\
\frac{1}{2} & \frac{1}{2} & F_t
\end{array}\right\} \left\{ \begin{array}{ccc}
F & F' & 1\\
J' & J & 0
\end{array}\right\} \left\{ \begin{array}{ccc}
\frac{1}{2} & J' & 1\\
J & \frac{1}{2} & 1
\end{array}\right\},
\\
\displaybreak[1]
\nonumber \\
\left\langle \Pi'\left|\frac{1}{r}e^{-r/\rho}\right|\Pi\right\rangle =  &\: \delta_{LL'}\delta_{JJ'}\delta_{F_{t}F_{t}'}\delta_{M_{F_{t}}M_{F_{t}}'}\:\frac{1}{4}N_{N'L}^{\left(1\right)}N_{NL}^{\left(1\right)}\:\sum_{k=0}^{N'}\:\sum_{j=0}^{N}\left(-1\right)^{k+j}\nonumber \\
\nonumber \\
\times &\: 
\binom{N'+2L+1}{N'-k}\binom{N+2L+1}{N-j}\frac{\left(2L+k+j+1\right)!}{k!\,j!}\left[\frac{2\rho}{1+2\rho}\right]^{\left(2+2L+k+j\right)}.
\end{align}

\end{widetext}


%

\end{document}